\documentclass[aps,prd,twocolumn,nofootinbib,superscriptaddress]{revtex4-1}
\usepackage[version=4]{mhchem}
\usepackage{url}
\usepackage{color}
\usepackage{graphicx}
\usepackage{epsfig}
\def\lsim{\:\raisebox{-0.5ex}{$\stackrel{\textstyle<}{\sim}$}\:}
\def\gsim{\:\raisebox{-0.5ex}{$\stackrel{\textstyle>}{\sim}$}\:}

\def\beq{\begin{equation}}
\def\eeq{\end{equation}}
\def\beqa{\begin{eqnarray}}
\def\eeqa{\end{eqnarray}}

\def\eminus{e^{-}}
\def\eplus{e^{+}}

\def\nue{\nu_e}
\def\anue{\bar{\nu}_e}
\def\numu{\nu_\mu}
\def\anumu{\bar{\nu}_\mu}
\def\nutau{\nu_\tau}
\def\anutau{\bar{\nu}_\tau}
\def\nui{\nu_i}

\def\Enu{E_\nu} 
\def\Enumin{E_{\nu}^{\rm min}} 
\def\Enui{E_{\nui}}

\def\Estar{E_*}
\def\En{E_n}
\def\NR{N_{\rm R}}
\def\ER{E_{\rm R}}
\def\Soneth{S1_{\rm th}} 
\def\Stwoth{S2_{\rm th}} 
\def\dsigmadER{\frac{d\sigma}{d\ER}}

\def\dNRdER{\frac{d\NR}{d\ER}}

\def\Ee{E_e}

\def\Enuiav{\langle\Enui\rangle}

\def\alphanui{\alpha_{\nui}}

\def\gammazero{\gamma_0}

\def\fm{\,{\rm fm}} 
\def\cm{\,{\rm cm}}

\def\musec{\,\mu{\rm s}}

\def\g{\,{\rm g}}

\def\erg{\,{\rm erg}}

\def\Msun{M_\odot}
\def\msun{M_\odot}

\newcommand{\CC}{\rm CC}

\def\cenns{CE$\nu$NS}
\def\nuein{$\nue$In}
\def\nueCC{$\nue$CC}
\def\nueigamma{$\nue$I$\gamma$}

\def\nueCCegamma{\nueCC\_($e^{-}+\gamma$)}
\def\nueCCe{\nueCC\_$e^{-}$}
\def\nueCCgamma{\nueCC\_$\gamma$}
\def\nueCCn{\nueCC\_$n$}

\def\sigmanueCC{\sigma_{\nue}^{\CC}}
\def\sigmanueCCn{\sigma_{\nue}^{{\CC},n}}

\newcommand{\GT}{\rm GT}
\newcommand{\F}{\rm F} 
\newcommand{\GTminus}{{\GT}_{-}}
\newcommand{\GTplus}{{\GT}_{+}}

\newcommand{\SGTminus}{S_{\GTminus}}
\newcommand{\SGTplus}{S_{\GTplus}}

\newcommand{\SF}{S_{\F}}
\def\Xe132{^{132}{\rm Xe}}
\def\Xeatom{^{132}_{\,\,\,54}{\rm Xe}}
\def\Cs132{^{132}{\rm Cs}} 
\def\Csatom{^{132}_{\,\,\,55}{\rm Cs}}
\def\I132{^{132}{\rm I}}

\begin{document}
\title{Inelastic charged current interaction of supernova neutrinos in 
two-phase liquid xenon dark matter detectors} 
\author{Pijushpani Bhattacharjee}
\email{pijush.bhattacharjee@saha.ac.in}
\affiliation{\mbox{Saha Institute of Nuclear Physics, HBNI, 1/AF 
Bidhannagar, Kolkata 700064, India}} 
\author{Abhijit Bandyopadhyay}
\email{abhijit@rkmvu.ac.in}
\affiliation{\mbox{Ramakrishna Mission Vivekananda Educational and 
Research Institute},\\
\mbox{Belur Math, Howrah 711202, India}}
\author{Sovan Chakraborty}
\email{sovan@iitg.ac.in}
\affiliation{\mbox{Department of Physics, Indian Institute of 
Technology - Guwahati, Guwahati 781039, India}}
\author{Sayan Ghosh}
\email{sayan.ghosh@saha.ac.in}
\affiliation{\mbox{Saha Institute of Nuclear Physics, HBNI, 1/AF 
Bidhannagar, Kolkata 700064, India}}
\author{Kamales Kar}
\email{kamales.kar@gm.rkmvu.ac.in}
\affiliation{\mbox{Ramakrishna Mission Vivekananda Educational and
Research Institute},\\
\mbox{Belur Math, Howrah 711202, India}}
\author{Satyajit Saha}
\email{satyajit.saha@saha.ac.in}
\affiliation{\mbox{Saha Institute of Nuclear Physics, HBNI, 1/AF
Bidhannagar, Kolkata 700064, India}}
\begin{abstract}
It has been known that neutrinos from supernova (SN) bursts can give 
rise to nuclear recoil (NR) signals arising from coherent elastic 
neutrino-nucleus scattering (\cenns) interaction, a neutral current (NC) 
process, of the neutrinos with xenon nuclei in future large (multi-ton 
scale) liquid 
xenon (LXe) detectors employed for dark matter search, depending on the 
SN progenitor mass and distance to the SN. In this paper, we show that 
the same detectors will also be sensitive to inelastic charged current 
(CC) interactions of the SN electron neutrinos (\nueCC) with the xenon 
nuclei. 
Such interactions, while creating an electron in the final state, also 
leave the post-interaction target nucleus in an excited state, the 
subsequent deexcitation of which produces, among other particles, gamma 
rays and neutrons. The electron and deexcitation gamma rays will give 
``electron recoil" (ER) type signals, while the deexcitation 
neutrons---the so called ``neutrino induced neutrons" 
(\nuein)---produce, through their multiple scattering on the xenon 
nuclei, further xenon nuclear recoils that will also give NR  
signals (in addition to those produced through the \cenns\  
interactions).   
We discuss the observable scintillation and ionization signals 
associated with SN neutrino induced \cenns\  and \nueCC\  events 
in a generic LXe detector and argue that 
upcoming sufficiently large LXe detectors should be able to detect both 
these types of events due to neutrinos from reasonably close by SN 
bursts. We also note that since the total CC induced 
ER and NR signals receive contributions predominantly 
from \nueCC\  interactions while the \cenns\  contribution comes from NC  
interactions of {\it all the six species of neutrinos}, identification 
of the \nueCC\  and \cenns\  origin events may offer the possibility of 
extracting useful information about the distribution of the total SN 
explosion energy going into different neutrino flavors. 

\end{abstract}
\maketitle

\section{Introduction}\label{sec:intro} 
Core collapse supernova (CCSN) explosions~\cite{SN-explosions} give out 
huge flux of neutrinos (and antineutrinos) of all 
flavors~\cite{scholberg-12,sovan-SN-nu-rev-16,Beacom-21} with energies 
up to a few tens of MeV over a time scale of $\sim$ 10 
seconds. These neutrinos carry almost all 
($\sim$ 99\%) of the gravitational energy ($\sim 10^{53}\erg$) released 
due to collapse of the core of the massive progenitor star. A  
number of large experimental facilities around the 
world~\cite{SN-nu-detectors} should be able to detect the neutrinos from 
the next nearby (hopefully Galactic) CCSN, after 
the historic first detection of neutrinos from the supernova 
SN1987A located in the Large Magellanic Cloud (LMC) at a distance of 
$\sim$ 50 kpc from Earth~\cite{SN1987A-detection}. Detection of  
supernova neutrinos of different flavors from a single supernova by 
multiple detectors has the potential to yield extremely valuable 
information about not only the supernova process itself but also  
various aspects of fundamental physics of neutrinos themselves.        

Neutrinos are detected through their weak charged current (CC) and 
neutral current (NC) interactions with electrons and nuclei. There is a 
large body of literature on possible neutrino interaction channels that 
can be employed for detection of SN neutrinos; see, e.g. 
Refs.~\cite{IBD} (inverse beta decay), \cite{nu-e-scatt} 
(neutrino-electron scattering), 
\cite{nu-nucleus,nu-nucl-scatt-revs} (neutrino-nucleus scattering) and  
\cite{cenns-original,Drukier-Stodolsky-cenns,Cabrera-etal-cenns,cenns-detection} 
(coherent elastic neutrino-nucleus scattering (\cenns)).  
Ref.~\cite{scholberg-12} gives 
a comprehensive review and references on various neutrino 
processes and their cross sections in the context of supernova neutrino 
detection. 

The NC process of \cenns\ has recently received much 
attention~\cite{cenns-dm-det-Horowitz,cenns-dm-det-Monroe,cenns-Anderson-11,sovan-pb-kk-14,lang-etal-16}
in the context of large (multi-ton scale) 
detectors~\cite{multi-ton-dm-detectors} searching for 
the weakly interacting massive particle (WIMP) candidates of dark 
matter (DM)~\cite{wimp-detection-review} through WIMP-induced nuclear 
recoils (NR). In \cenns, neutrinos 
undergo coherent elastic scattering on the target nucleus with a cross 
section that is enhanced by the square of the number of neutrons in the 
target nucleus. The coherent enhancement of the cross section is of  
particular interest for SN neutrinos because of their relatively low 
energies (a few to a few tens of MeV) where the coherence effect 
is dominant. The recoiling target 
nucleus of mass $M$ gets a maximum kinetic energy of $2{\Enu}^2/M$, 
where $\Enu$ is the neutrino energy. There is a trade-off between the 
cross section enhancement (requiring 
large mass nuclei) and the maximum recoil energy ($\propto 1/M$), with 
the latter typically in the region of a few keV for neutrino 
energy of $\sim 10$ MeV and target nuclei with mass number in the 
region of $\sim 100$ desired for reasonably large cross section 
(typically, $O(10^{-39}\cm^2)$). Such low recoil 
energies, while very difficult to detect in conventional neutrino 
detectors, are, however, within reach 
of the large WIMP DM detectors. Thus, sufficiently large WIMP 
DM detectors with suitably chosen detector materials can also be 
sensitive to neutrinos from individual SN events. Importantly, the 
\cenns, it being a NC process, is equally sensitive to all flavors of 
neutrinos (and antineutrinos), which offers a probe for estimating the 
total explosion energy going into 
neutrinos~\cite{Drukier-Stodolsky-cenns,cenns-dm-det-Horowitz}, and also 
possibly for demarcating the neutrinos originating in the different 
temporal phases of the neutrino emission during the SN 
event~\cite{sovan-pb-kk-14}. Because of these reasons, detailed 
calculations have been done studying the \cenns\  sensitivity of the 
next generation WIMP DM detectors, in particular, those using LXe as 
target detector material, to possible future nearby (Galactic) 
SN events; see, for example, Ref.~\cite{lang-etal-16}. 

In the present paper we show that, in 
addition to the NC process of 
\cenns, the future large LXe DM detectors would also be 
sensitive to inelastic CC interactions of the SN electron neutrinos 
(\nueCC) with the xenon nuclei in the LXe detectors. 
Such interactions would produce a final state electron carrying a 
substantial fraction of the incident neutrino energy, and at the same 
time leave the post-interaction target nucleus in an 
excited state. Subsequent deexcitation of the latter would produce, 
among other particles, gamma rays and 
neutrons~\cite{nuin-process-main,Kolbe-Langanke-01,Engel-etal-03,Bandyo-etal-17,bhatta-kar-EPJST-21}.  
The electron and deexcitation gamma rays would give rise to signals 
similar to those seen for the ``electron recoil" (ER) events 
~\cite{Xenon1t-16} in DM detectors. These classes of signals will be  
designated as \nueCCe\  and \nueCCgamma\   throughout this paper. At 
the same time, the ``{\it neutrino induced neutrons}" (\nuein) produced 
in the deexcitation of the post-interaction target nucleus would undergo 
elastic scattering on the target xenon nuclei and give rise to 
additional xenon nuclear recoils that, as we show below, may give 
significant contribution to observable NR signals at relatively large 
($\gsim$ 30 keV) recoil energies where the \cenns\  
generated NR signal contribution is relatively small. In this paper we 
shall denote these \nuein\  generated NR signals by \nueCCn. 

Below we discuss the observable scintillation and ionization signals
associated with the SN neutrino induced \cenns\  and \nueCC\  events
in a generic two-phase xenon detector using the GEANT4~\cite{GEANT4} 
simulation toolkit, and study their detectability in future multi-ton 
scale LXe based DM detectors \cite{multi-ton-dm-detectors}. 
For simplicity, in this paper we consider only a single xenon isotope, 
namely, $\Xe132$, for our calculations in order to illustrate the 
basic physical processes involved. 

While both $\nue$s and $\anue$s from the SN can undergo CC 
interactions with the xenon nuclei (the $\numu$s, $\anumu$s, 
$\nutau$s and $\anutau$s can undergo only NC interaction because of 
insufficient incoming neutrino energies for production of the associated 
charged leptons), the contribution from $\anue$ CC interaction with 
$\Xe132$, which involves conversion of protons to neutrons inside the 
nucleus, is strongly suppressed due to Pauli blocking of the neutron 
single particle states in the neutron rich final state 
nucleus~\cite{nu-nucl-scatt-revs}. Similarly, as seen in 
the case of $^{208}{\rm Pb}$ \cite{Bandyo-etal-17}, the contribution 
from the NC interaction of neutrinos and antineutrinos of all flavors 
with the target nuclei---in which the incoming neutrino or 
antineutrino only imparts energy to the neutrons and protons inside 
the nucleus without inducing inter-conversion between the nucleons---is 
also expected to be subdominant to the contribution from CC 
interactions of $\nue$s for a target nucleus like $\Xe132$ with a 
moderately large neutron excess ($N-Z = 24$), again partly due to Pauli 
blocking effects. In the present paper, therefore, we only consider the 
\nueCC\  interaction, i.e., the process 
$\Xe132\,(\nue,\eminus){\Cs132}^*$, for simplicity. 

For the SN neutrino flux to be used for numerical calculations in 
this paper, we use the results of the Basel-Darmstadt (BD) 
simulations~\cite{BD-SN-sim-10} of a $18\Msun$ progenitor star placed at 
a distance of 1 kpc from earth for illustration of our 
results~\cite{fnote-sn-distance}. The instantaneous energy spectrum, 
SN neutrinos of type $\nui$ (with $\nui \equiv \nue\,, \anue\,, \numu\,,
\anumu\,, \nutau\,,\anutau\,$) emitted (at the neutrinosphere)
per unit time per unit energy is usually 
parametrized in terms of the time-dependent neutrino 
luminosities ($L_{\nui} (t)$), average energies ($\Enuiav (t)$), and the 
spectral shape parameter $\alphanui (t)$ of the different neutrino 
flavors~\cite{Keil-etal-03}. The temporal profiles of these parameters 
for different neutrino flavors extracted from the BD simulation results 
for the $18\Msun$ progenitor SN used in this paper are given in 
graphical form in 
Refs.~\cite{sovan-pb-kk-14,Bandyo-etal-17,Borriello-Sovan-etal-12}. 
Note that the $\numu\,, \anumu\,, \nutau\,,\anutau\,$ have 
essentially identical spectra and temporal profiles of their 
luminosities and average energies in the BD simulations.  

For the \cenns\  process, since it is equally sensitive to all 
the neutrino species, the flavor oscillation effects are not relevant, 
and the time-integrated (over the duration of the SN
burst event) sum of fluxes of all the neutrino species 
reaching the earth is same as the sum of the number of neutrinos of all 
species emitted (at the neutrinospheres) per unit time and energy,  
divided by $4\pi d^2$, $d$ being the distance to the SN.    
On the other hand, the flux of $\nue$s would in general depend 
on various flavor oscillation processes, including the standard 
Mikheyev-Smirnov-Wolfenstein (MSW) matter-enhanced 
neutrino flavor oscillation (see, e.g., \cite{Mohapatra-Pal-book-2004}) 
as well as the phenomenon of collective neutrino flavor oscillations 
due to $\nu$-$\nu$ interaction in the deep interior of the SN progenitor 
star; see, e.g., 
Refs.~\cite{sovan-SN-nu-rev-16,Duan-etal-10,sovan-etal-NPB-16,Capozzi-Dasgupta-etal-19}. 
However, as discussed in 
Refs.~\cite{Borriello-Sovan-etal-12,sovan-etal-11}, for the 
time-integrated fluxes of different neutrino species, the collective 
oscillation effects are small, and the fluxes of various neutrino 
species reaching the earth are given essentially by the MSW oscillations 
in the supernova matter. The resulting expressions for the $\nue$ 
and $\anue$ fluxes at earth, for normal ordering (NO) and inverted 
ordering (IO) of the neutrino mass hierarchy are given in 
Refs.~\cite{Bandyo-etal-17,Borriello-Sovan-etal-12}, which we use in 
this paper. For simplicity, in this paper, we show all our numerical 
results for the NO case of neutrino mass hierarchy only; similar 
calculations can be done for the IO case. 

The rest of the paper is arranged as follows: 
In Sec.~\ref{sec:cenns} we set up the equations giving the 
differential recoil spectrum due to \cenns\  interaction of the SN 
neutrinos with the target nucleus. In Sec.~\ref{sec:nue-Xe132-CC} 
we discuss the inelastic CC interaction of $\nue$s with $\Xe132$ 
nuclei, and calculate the energy spectrum of the electrons produced in 
the primary CC process $\Xe132(\nue\,, \eminus){\Cs132}^*$, as well as 
the energy spectra of the neutrons and $\gamma$-rays
from deexcitation of the ${\Cs132}^*$ nuclei, for a $18\msun$ SN at a 
distance of 1 kpc from earth. The $\Xe132$ recoil spectrum due to 
multiple scattering of the neutrino-induced neutrons with the xenon 
nuclei is then calculated in Sec.~\ref{sec:nuin-Xe132-recoil-spect} 
using GEANT4 simulation. Sec.~\ref{sec:S1-S2} discusses the observable 
scintillation and ionization signals due to (a) $\Xe132$ nuclear recoils 
originating from \cenns\  interactions and (b) \nueCC\   
origin electrons, $\gamma$-rays and neutrons (with the last one 
giving $\Xe132$ nuclear recoils) in a generic dual-phase LXe detector. 
The contributions of the different signal components associated with the 
\cenns\  and \nueCC\  events are assessed and the possibility of 
distinguishing amongst them is discussed in 
Sec.~\ref{sec:event-type-contributions}. 
Finally, summary and conclusions are presented 
in Sec.~\ref{sec:summary}. 

\section{Supernova neutrino induced xenon recoil spectrum due to \cenns}
\label{sec:cenns}
The differential cross section, $\dsigmadER$, for the \cenns\  
interaction 
of a neutrino of energy $\Enu$ with a target nucleus of mass $M$ which 
is left with a recoil (kinetic) energy $\ER$ after the interaction  
is given 
by~\cite{cenns-original,Drukier-Stodolsky-cenns,cenns-Anderson-11}  
\beq
\dsigmadER = \frac{G_F^2 M}{4\pi}Q_W^2 \left(1-\frac{M\ER}{2\Enu^2}   
\right) F^2(\ER)\,,
\label{eq:cenns-dsigmadER}
\eeq 
where $G_F$ is the Fermi constant,   
$Q_W=N-(1-4\sin^2\theta_W)Z$ is the weak nuclear hypercharge for the  
nucleus with $N$ neutrons and $Z$ 
protons, $\theta_W$ is the weak mixing angle (with $\sin^2\theta_W\simeq 
0.2386$), and $F(\ER)$ is the nuclear form factor which we shall take to 
be of the Helm form~\cite{Lewin-Smith-96},  
\beq
F(\ER)=\frac{3 j_1(qr_n)}{qr_n}e^{-q^2s^2/2}\,, 
\label{eq:Form-factor}
\eeq 
where $j_1$ is the spherical Bessel function, $q=(2M\ER)^{1/2}$ is  
the momentum transfer to the nucleus, 
$s\simeq 0.9\fm$ is the nuclear skin thickness, 
$r_n^2=c^2+\frac{7}{3}\pi^2a^2-5s^2$ is the square of the effective 
nuclear radius with $c\simeq 1.23 A^{1/3} - 0.60\fm$, $a\simeq 
0.52\fm$, and $A=N+Z$. 

The differential recoil spectrum, $\dNRdER$, per ton (1000 kg) of target 
detector material (liquid $\Xe132$, in our case) for the 
time-integrated (over the duration of the SN event) neutrino flux is 
then given by 
\beq
\dNRdER = N_{\rm Xe} \sum_{\nui} \int_{\Enumin} d\Enu\,\dsigmadER\,  
\Phi_{\nui}(\Enu)\,,
\label{eq:dNRdER}   
\eeq
where the $\nui$ sum is over all six species of neutrinos, 
$\Enumin=(M\ER/2)^{1/2}$ is the minimum energy of a neutrino that can 
produce a recoiling $\Xe132$ nucleus of recoil energy $\ER$, $M$ being 
the mass of a $\Xe132$ nucleus, $N_{\rm Xe}=4.56\times 10^{27}$ is the 
number of $\Xe132$ nuclei in one ton of liquid $\Xe132$, and 
$\Phi_{\nui}(\Enu)$ is the time-integrated flux of SN $\nui$s per unit 
energy and per unit area at earth. 

To facilitate easy comparison, the resulting $\Xe132$ recoil spectrum 
due to \cenns\ process for the BD $18\Msun$ SN 
is displayed and discussed in section 
\ref{sec:nuin-Xe132-recoil-spect} together with the recoil spectra due 
to the neutrons coming from the CC \nuein\  process discussed in the 
next section. 

\section{Inelastic charged current interaction of supernova neutrinos 
with $\Xe132$ nuclei}
\label{sec:nue-Xe132-CC}
The inelastic CC interaction of a supernova $\nue$  
with a $\Xe132$ nucleus produces an electron and a $\Cs132$ nucleus in 
an excited state:  
\beq
\nue + \Xeatom \rightarrow \eminus + {\Csatom}^*\,,
\label{eq:nue-xe-CC-reaction} 
\eeq
where the superscript $*$ stands for the excited state of the 
final state nucleus.  

The differential cross section for this reaction in the $q\to 0$ 
limit (applicable for the energy range of SN neutrinos, $q$ being 
the momentum transfer) can be written 
as~\cite{Kuramoto-etal-90,Fuller-etal-99,Kolbe-etal-99}
\beqa
\frac{d\sigmanueCC}{d\Estar}(\Enu,\Estar) =
\frac{G_F^2 \cos^2\theta_c}{\pi} \,\,\, p_e E_e F(Z+1,E_e)\nonumber\\
  \hskip-1cm \times \left[\SF(\Estar) +
(g^{\rm eff}_{\rm A})^2 \SGTminus(\Estar)\right]\,,
\label{eq:nue_CC_diff_xsec}
\eeqa
where $G_F$ is the Fermi constant, $\theta_c$ is the Cabibbo angle,
$\Estar=\Enu-\Ee-Q$ is the excitation energy of the $\Cs132$ nucleus 
with $Q=M(\Csatom)-M(\Xeatom)=2.1263$ MeV~\cite{Q-value}, $M(\Csatom)$ 
and $M(\Xeatom)$ being the masses of the two nuclei,    
$\Enu$ is the incoming neutrino energy, and $\Ee$ and $p_e$ are the  
energy and momentum of the emitted electron, respectively.  
The quantities $\SF(\Estar)$ and $\SGTminus (\Estar)$ are, respectively, 
the modulus squares of the averaged Fermi and Gamow-Teller ($\GTminus$) 
transition matrix elements between the ground state of the initial 
nucleus $\Xeatom$ and the excited state of the final nucleus 
${\Csatom}^*$, 
and $g^{\rm eff}_{\rm A}\simeq 1.26$ is the ratio of the effective axial 
vector to vector coupling constants of the bare nucleon in the $q\to 0$ 
limit. The general expressions for $\SF(\Estar)$ and $\SGTminus 
(\Estar)$ are given  
in Refs.~\cite{Kuramoto-etal-90,Fuller-etal-99,Kolbe-etal-99}.   

The factor $F(Z+1,E_e)$ in Eq.~(\ref{eq:nue_CC_diff_xsec}) is the 
correction factor which accounts 
for the distortion of the outgoing electron wave function due to 
Coulomb interaction with the final nucleus $(N-1,Z+1)$ (with the initial 
nucleus represented as $(N,Z)$), and is given by~\cite{Engel-98}
\beq
F(Z,\Ee) = 2 (1+\gammazero)(2p_e R)^{2(\gammazero-1)}
\times\,\, \frac{\arrowvert\Gamma(\gammazero+iy)\arrowvert^2}{
\arrowvert\Gamma(2\gammazero+1)\arrowvert^2}\exp(\pi y)\,,
\label{eq:FermiFunction}
\eeq
where $\gammazero=\left(1-Z^2\alpha^2\right)^{1/2}$, $y=\alpha Z 
E_e/p_e$, $R$ is the radius of final nucleus and $\alpha$ the fine 
structure constant.

The Fermi transitions have the isospin selection rule 
$\Delta T=T^{'}-T=0$ where $T^{'}$ and $T$ are, respectively, the
isospins of the final and initial nucleus. Also, the Fermi strength sum 
over the final nucleus states is equal to $(N-Z)$, and it goes almost 
completely to the Isobaric Analog state (IAS) of the final nucleus with 
a very small spread in the neighboring states due to isospin breaking 
essentially by Coulomb interaction. The IAS in ${\Cs132}^*$ is at an 
excitation of 13.8 MeV~\cite{Unlu-16}. On the other hand, the 
Gamow-Teller strength distribution is broad and its strength sum, 
which is slightly larger than $3(N-Z)$, is distributed over states with 
$T^{'} = T-1, T$ and $T+1$, i.e., with $\Delta T=1$. For the GT strength 
distribution $\SGTminus$ for the final nucleus ${\Cs132}^*$ we use the 
results of a theoretical calculation~\cite{Moreno-06} done within the 
deformed Hartree-Fock formalism using the density dependent Skyrme 
interaction Sk3~\cite{Beiner-75}. 

The reaction $\Xe132(\anue\,, \eplus){\I132}^*$ initiated by the 
supernova 
$\anue$s also takes place, but the total GT strength, $\SGTplus$, for 
the reaction is $\simeq 0.51$~\cite{Moreno-06}, 
which is only 0.7\% of the total $\GTminus$ strength of $\simeq 
72.12$~\cite{Moreno-06} for the 
reaction $\Xe132(\nue\,, \eminus){\Cs132}^*$ discussed above.  
Hence, in this paper we neglect the $\anue$ CC contribution compared to 
the contribution from $\nue$ capture reaction.     

\begin{figure}
\includegraphics[width=0.9\columnwidth]{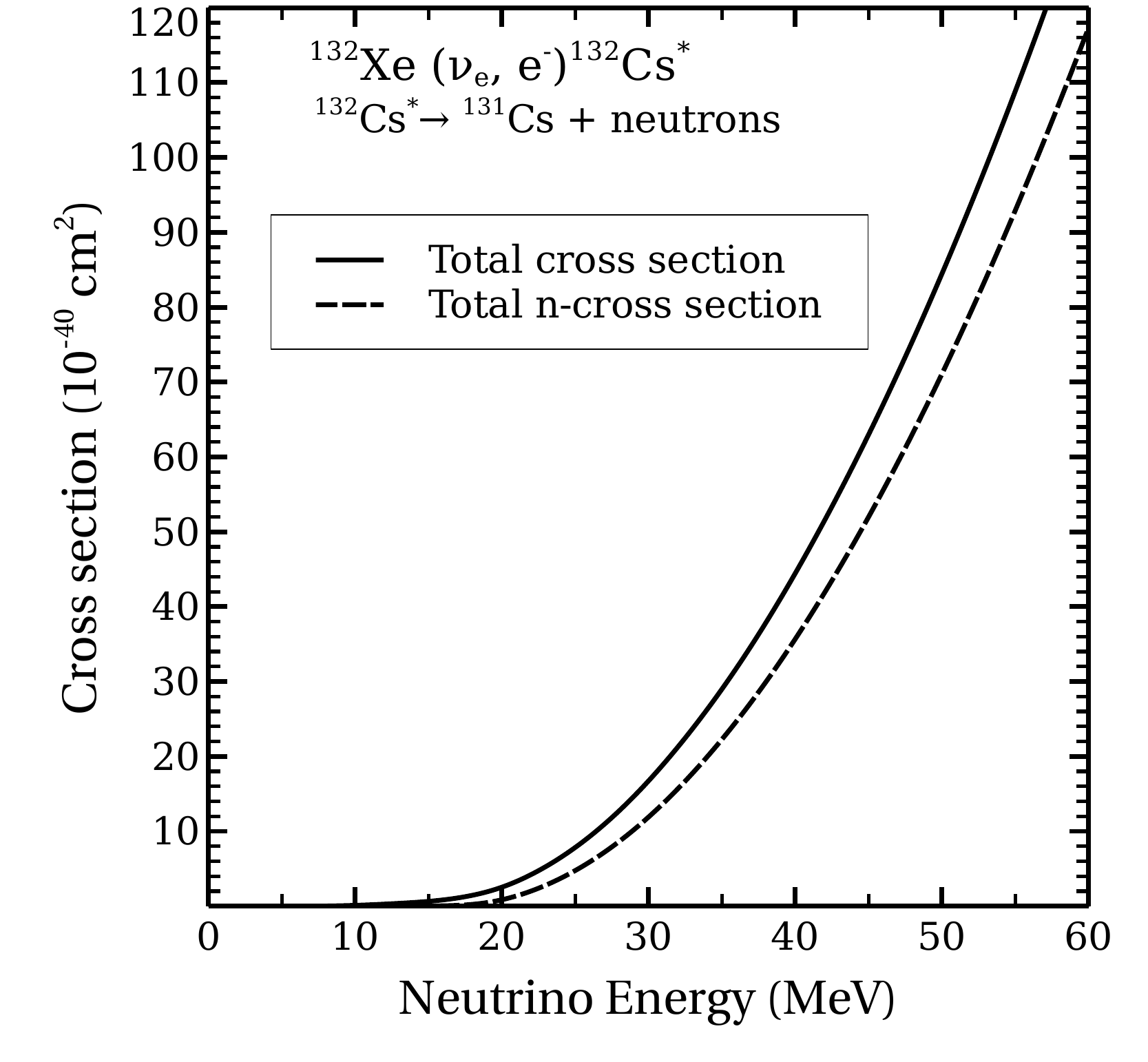}
\caption{Total cross section (solid curve) and neutron emission cross 
section (dashed curve) for 
the CC interaction of $\nue$ with $\Xe132$ as functions of the incoming 
neutrino energy.}
\label{fig:total-and-n-xsecs}
\end{figure}

The total cross section for \nueCC\   interaction of a neutrino of
energy $\Enu$ with $\Xe132$ is
obtained by integrating Eq.~(\ref{eq:nue_CC_diff_xsec}) over
$\Estar$, and is shown in Fig.~\ref{fig:total-and-n-xsecs} as a function 
of the incoming neutrino energy. 

\subsection{Electron energy spectrum}
\label{subsec:e-spect}
\begin{figure}
\includegraphics[width=0.9\columnwidth]{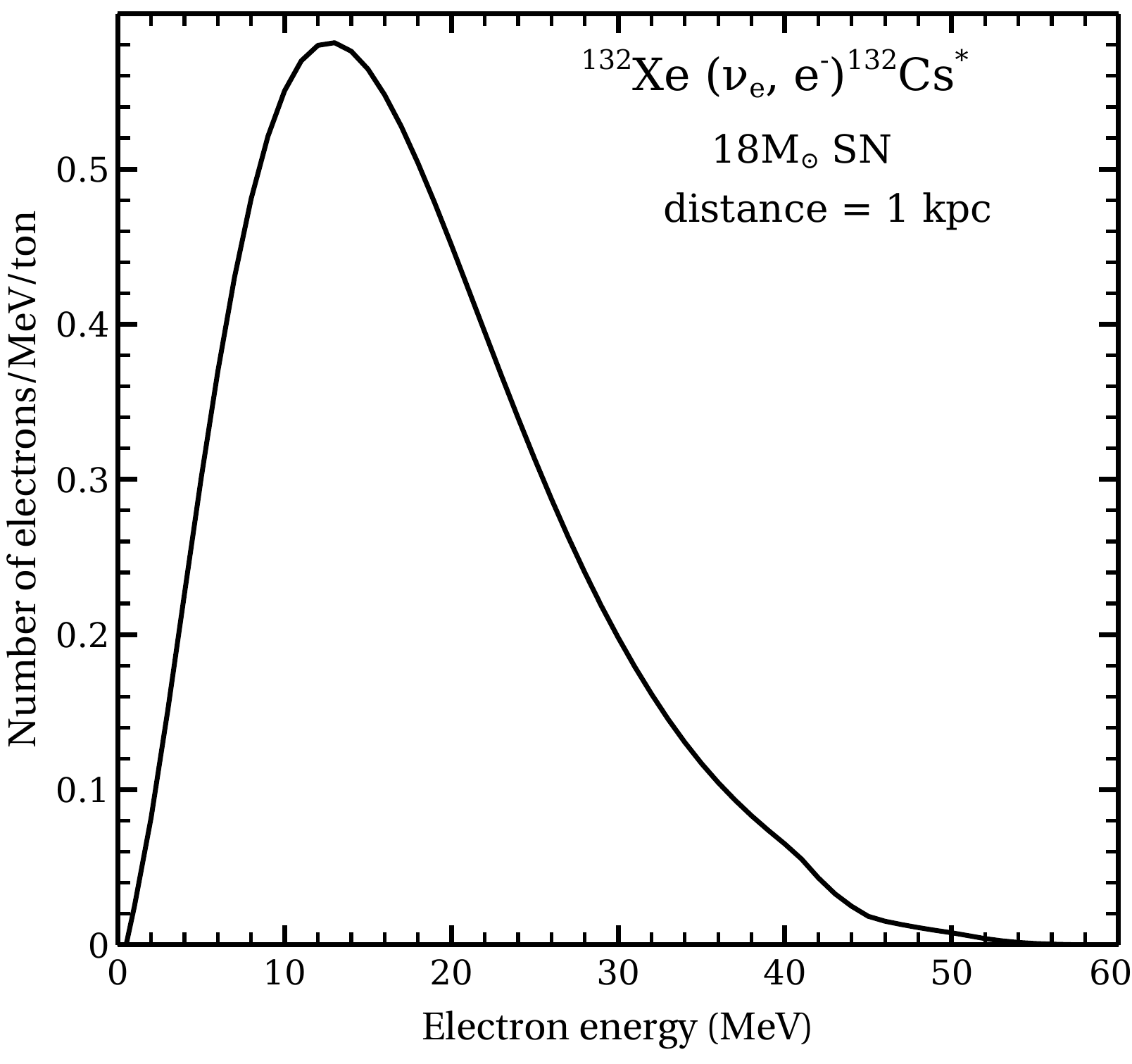}
\caption{Energy spectrum of electrons produced due to CC interaction of 
supernova $\nue$s with $\Xe132$ nuclei, for the 
case of normal ordering (NO) of neutrino mass 
hierarchy, for SN neutrino flux given by the Basel-Darmstadt 
simulations~\cite{BD-SN-sim-10} of a $18\msun$ progenitor supernova at a 
distance of 1 kpc.}
\label{fig:e-spectrum-NO}
\end{figure}

The total energy spectrum of the electrons produced by the incident
flux of SN $\nue$s per ton of $\Xe132$ is given by
\beq
\frac{dN_{e,{\rm total}}^{\CC}}{d\Ee}=N_{\rm Xe} \int d\Enu
\Phi_{\nue}(\Enu)
 \frac{d\sigmanueCC}{d\Ee}(\Enu,\Ee)\,,
\label{eq:total-e-spect}
\eeq
where $\Phi_{\nue}(\Enu)$ is the time-integrated flux (number
per unit area per unit energy) of the SN $\nue$s falling on the
detector, $N_{\rm Xe}=4.56\times 10^{27}$ is the number of $\Xe132$ 
nuclei in one ton of liquid $\Xe132$, 
and $\frac{d\sigmanueCC}{d\Ee}(\Enu,\Ee)$ is given by 
Eq.~(\ref{eq:nue_CC_diff_xsec}) with $\Estar=\Enu-\Ee-Q$. The 
resulting electron energy spectrum for the BD $18\Msun$ SN $\nue$ flux
is shown in 
Fig.~\ref{fig:e-spectrum-NO}. The total number of 
electrons produced is $\sim$ 13.0. 

\subsection{Energy spectrum of the neutrino induced neutrons (\nuein)}
\label{subsec:nuin-spect}
The final state excited nucleus ${\Cs132}^*$ can decay by emitting one 
or more neutrons: 
\beq
{\Cs132}^*\, \rightarrow\, {^{131}{\rm Cs}} + n\, \quad 
{\mathrm{or}}\quad 
{\Cs132}^*\, \rightarrow\, {^{130}{\rm Cs}} + 2n\,,
\eeq
and so on. 

The emission of neutrons by the neutrino-induced excited nucleus can be 
considered as a two-step process involving two independent physical 
processes~\cite{Kolbe-Langanke-01,Engel-etal-03}: the production of a 
final state excited nucleus due to absorption of the incoming neutrino 
by the target nucleus in the first step, and subsequent deexcitation of 
the final state nucleus through neutron emission in the second step.

The differential cross section for the first step, i.e., 
production of the excited $\Cs132$ nucleus due to $\nue$ capture on a 
target $\Xe132$ nucleus is given by Eq.~(\ref{eq:nue_CC_diff_xsec}). 
With this, the energy spectrum of the emitted neutrons is calculated in
the following way: The differential cross section for emission of
neutrons per unit neutron (kinetic) energy $\En$ by a single 
final state nucleus due to inelastic CC interaction of an incoming 
$\nue$ of energy $\Enu$ with the target nucleus can be written 
as 
\beq
\frac{d\sigmanueCCn}{d\En}(\Enu,\En)
= \int \frac{d\sigmanueCC}{d\Estar}(\Enu,\Estar)
\frac{dN_n}{d\En}(\Estar,\En)\, d\Estar\,,
\label{eq:dsigman_dEn}
\eeq
where $\frac{dN_n}{d\En}(\Estar,\En)$ is the energy spectrum of the
neutrons produced by the excited nucleus of excitation energy $\Estar$. 
\begin{figure}
\includegraphics[width=0.9\columnwidth]{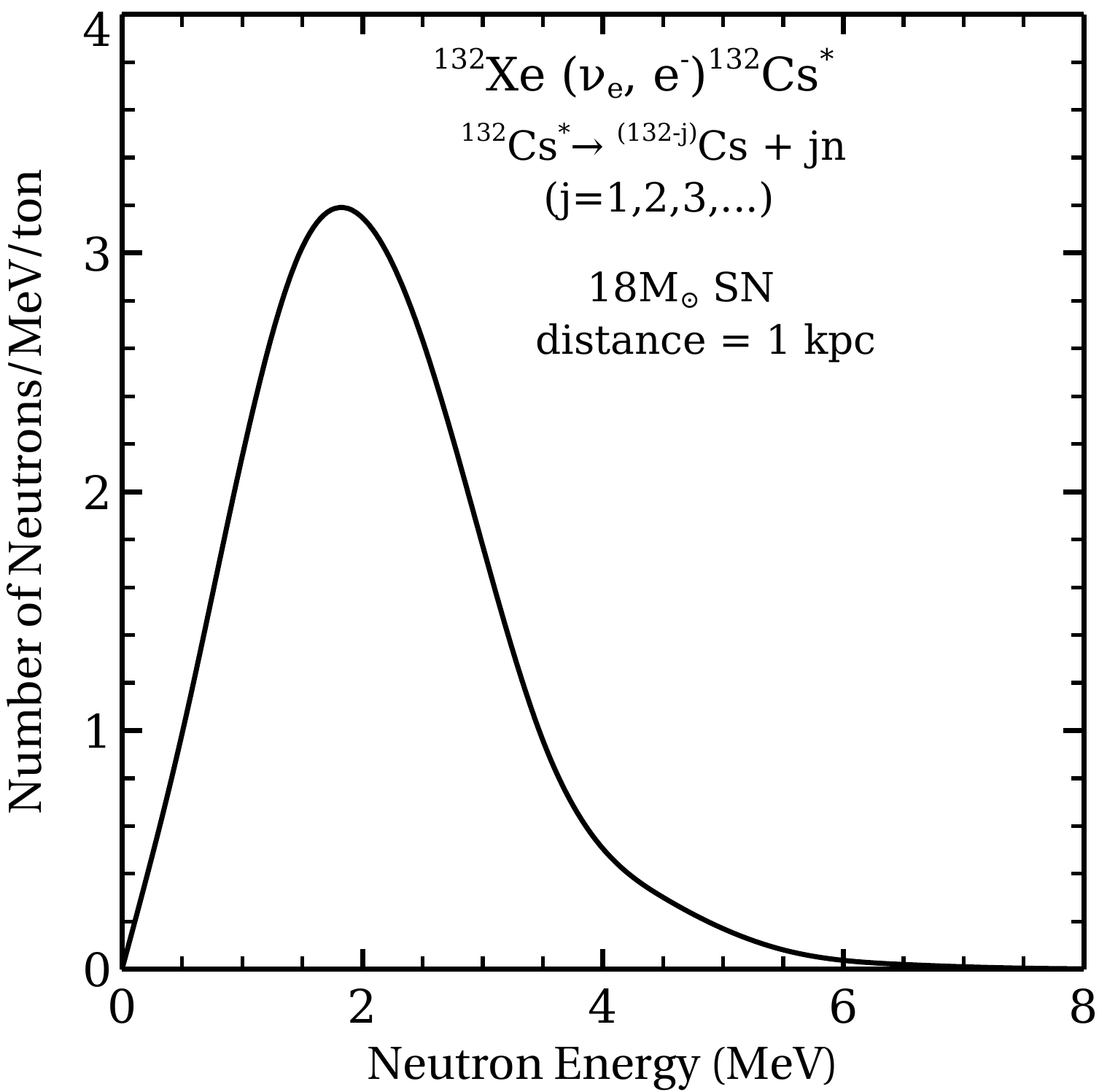}
\caption{Energy spectrum of neutrons emitted by excited $\Cs132$ nuclei 
produced due to inelastic CC interaction of supernova $\nue$s 
with $\Xe132$ nuclei, for the 
case of normal ordering (NO) 
of neutrino mass hierarchy, for SN neutrino 
flux given by the Basel-Darmstadt simulations~\cite{BD-SN-sim-10} of a 
$18\msun$ progenitor supernova at a distance of 1 kpc.}
\label{fig:n-spectrum-NO}
\end{figure}

We calculate the neutron energy spectrum, 
$\frac{dN_n}{d\En}(\Estar,\En)$, for $\Cs132^{*}$ using the 
fusion-evaporation Monte Carlo code PACE4~\cite{PACE4-code} originally 
developed by Gavron~\cite{Gavron-80}, including contributions 
from 1-, 2- and 3-neutron emissions. The total neutron emission cross 
section obtained by integrating Eq.~(\ref{eq:dsigman_dEn}) 
over $\En$ is shown in Fig.~\ref{fig:total-and-n-xsecs} as a function  
of the incoming neutrino energy, which can be compared with the total 
$\nue$ CC cross section shown in the same Figure. 

The total energy spectrum of the neutrons produced by the incident
flux of SN neutrinos is then given by
\beq
\frac{dN_{n,{\rm total}}^{\CC}}{d\En}=N_{\rm Xe} \int d\Enu
\Phi_{\nue}(\Enu)
 \frac{d\sigmanueCCn}{d\En}(\Enu,\En)\,,
\label{eq:total-n-spect}
\eeq
where $\Phi_{\nue}(\Enu)$ is the time-integrated flux spectrum (number
per unit area per unit energy) of the SN $\nue$s falling on the
detector. The resulting neutron energy spectrum for the BD $18\Msun$ SN 
$\nue$ flux 
for the NO case of neutrino mass hierarchy
is shown in 
Fig.~\ref{fig:n-spectrum-NO}. 
The total number of neutrons 
produced is $\sim$ 8.0. 

\subsection{Energy spectrum of the neutrino induced $\gamma$-rays  
(\nueigamma)}
\label{subsec:nuigamma-spect}
\begin{figure}
\includegraphics[width=0.9\columnwidth]{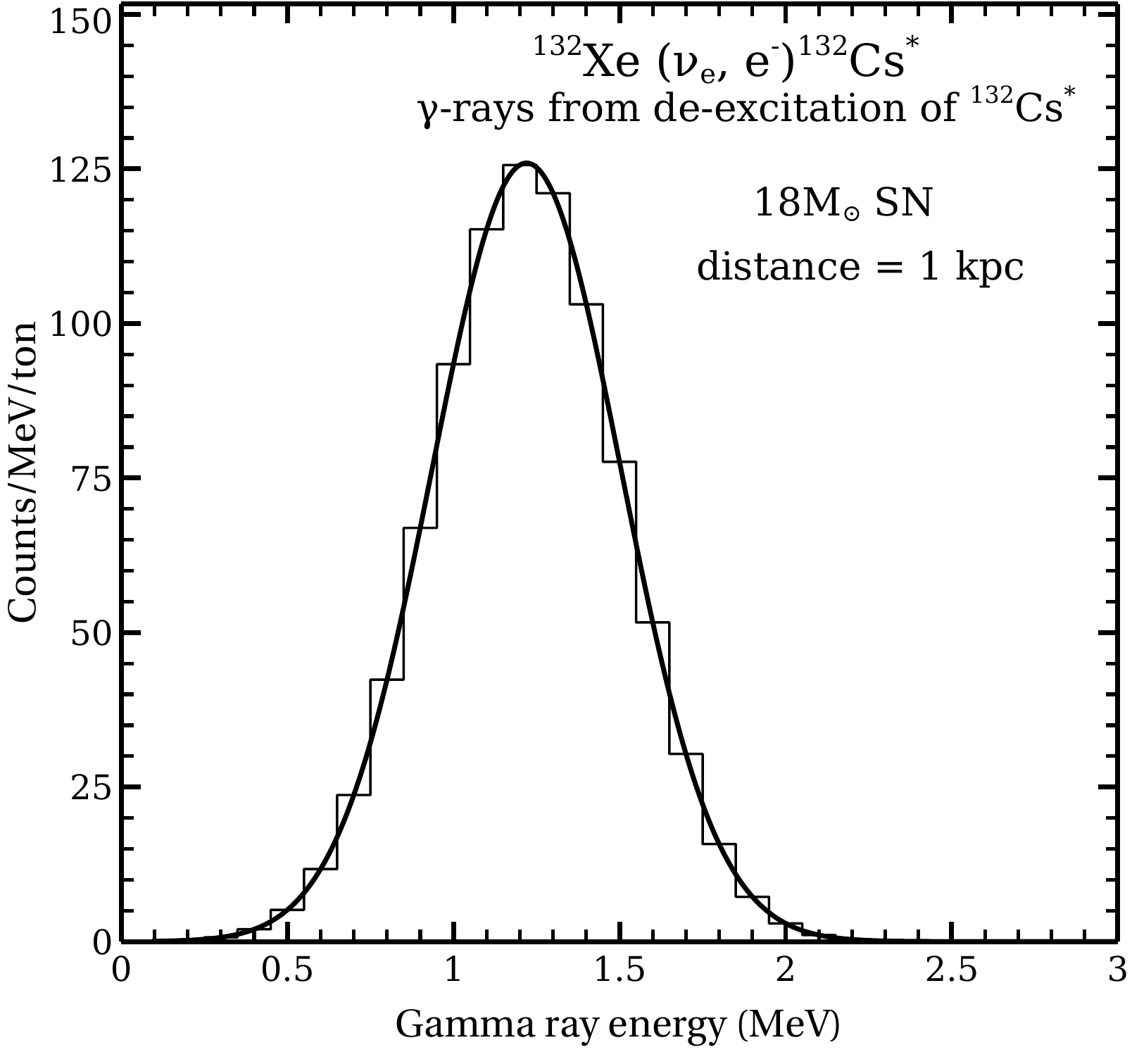}
\caption{Energy spectrum of deexcitation $\gamma$-rays emitted by 
excited $\Cs132$ 
nuclei produced due to inelastic CC interaction of supernova $\nue$s
with $\Xe132$ nuclei, for the
case of normal ordering (NO)
of neutrino mass hierarchy, for SN neutrino
flux given by the Basel-Darmstadt simulations~\cite{BD-SN-sim-10} of a
$18\msun$ progenitor supernova at a distance of 1 kpc.}
\label{fig:gamma-spectrum-NO}
\end{figure}
The final state excited nucleus ${\Cs132}^*$ can also decay by emitting 
$\gamma$-rays. These gamma rays are of two
types: discrete gamma rays arising from decays from higher energy 
bound states of the nucleus (${\Cs132}^*$ in the present case) to lower 
energy ones, and statistical gamma rays having a continuous spectrum. 
The gamma rays of the latter type dominate and compete with
the emission of neutrons and other light charged particles. 
While the discrete gamma rays predominantly have energy below 1 MeV, a 
calculation using the PACE code~\cite{PACE4-code} gives an 
yield of continuum $\gamma$-rays that are concentrated in the energy 
range of $\sim$ 1--2 MeV. Ignoring the discrete gamma rays, the average 
energy and average multiplicity of the deexcitation $\gamma$-rays from 
${\Cs132}^*$ nuclei are found to be 
$\sim$ 1.2 MeV and $\sim$ 6 per ton of liquid $\Xe132$, respectively, 
with a total number of $\gamma$-ray photons of $\sim$90, for the same SN 
(mass $18\Msun$) at a distance of 1 kpc as used in the calculation for 
neutron emission, for the case of normal ordering of neutrino mass 
hierarchy. The resulting spectrum of $\gamma$-rays is closely 
approximated by a gaussian with a mean at 1.22 MeV and FWHM of 0.67 MeV, 
and is shown in Fig.~\ref{fig:gamma-spectrum-NO}.

Inclusion of discrete gamma rays would slightly lower the mean energy, 
but this is effectively taken care of by increasing the 
width of the Gaussian distribution. On the other hand, the higher 
energy ($>$ 2 MeV) part of the spectrum, which sums up to only $\lsim 
0.1$\% of the contribution included in the spectrum, rapidly falls
off with energy. 

\section{Xenon recoil spectrum due to supernova neutrino induced
neutrons} 
\label{sec:nuin-Xe132-recoil-spect}
The neutrons produced by inelastic CC interaction of the supernova 
$\nue$s with target $\Xe132$ nuclei will scatter off the xenon nuclei 
themselves giving rise to recoiling $\Xe132$ nuclei. From 
Fig.~\ref{fig:n-spectrum-NO} we see that the neutrons produced by the 
supernova neutrinos typically have energies of a few MeV with a spectrum 
peaking at $\sim$ 2 MeV. At these neutron energies, elastic 
scattering dominates~\cite{ENDF}. The mean-free-path (m.f.p) of neutrons 
of energies in the range of a few 100 keV to a few MeV in LXe  
is $O(10)$ cm. So, depending on the dimensions of the active volume of 
LXe and the point of production of the neutron within the 
detector, a neutron of a given initial energy may undergo multiple 
scattering within the detector giving rise to multiple xenon recoils 
with different recoil energies. To take this into account we perform a 
simulation of neutron scattering on xenon 
nuclei~\cite{Sayan-TAUP-paper-2022} using the GEANT4 
simulation toolkit~\cite{GEANT4}. We use the GEANT4 
Physics List {\it QGSP\_BERT\_HP} \  which, in addition to including 
all the 
standard electromagnetic processes, employs the high precision 
neutron package ({\it NeutronHP}) to transport neutrons below 20 MeV 
down to thermal energies~\cite{GEANT4-neutron-physics-list}.  

\subsection{GEANT4 simulation of neutron scattering on xenon nuclei}
\label{subsec:geant4}
Guided by the typical designs of the LXe detectors being used 
in dark matter search experiments~\cite{multi-ton-dm-detectors}, we 
consider 1 ton of liquid $\Xe132$ contained in a cylindrical volume 
of aspect ratio (diameter-to-height) of 1:1. With a density of $\sim 
2.953 \g/\cm^3$ of liquid xenon, both the diameter and height of the 
cylinder are taken to be $\sim$ 75.4 cm. As far as the 
neutron induced xenon recoils are concerned, we find no significant 
differences in the results for the recoil spectra whether we consider 
natural xenon or one of its isotope (here $\Xe132$). Neutrons of a  
given initial energy are generated homogeneously within the LXe  
volume with randomly chosen initial positions and with isotropic initial 
velocity directions.  

\begin{figure}
\includegraphics[width=\columnwidth]{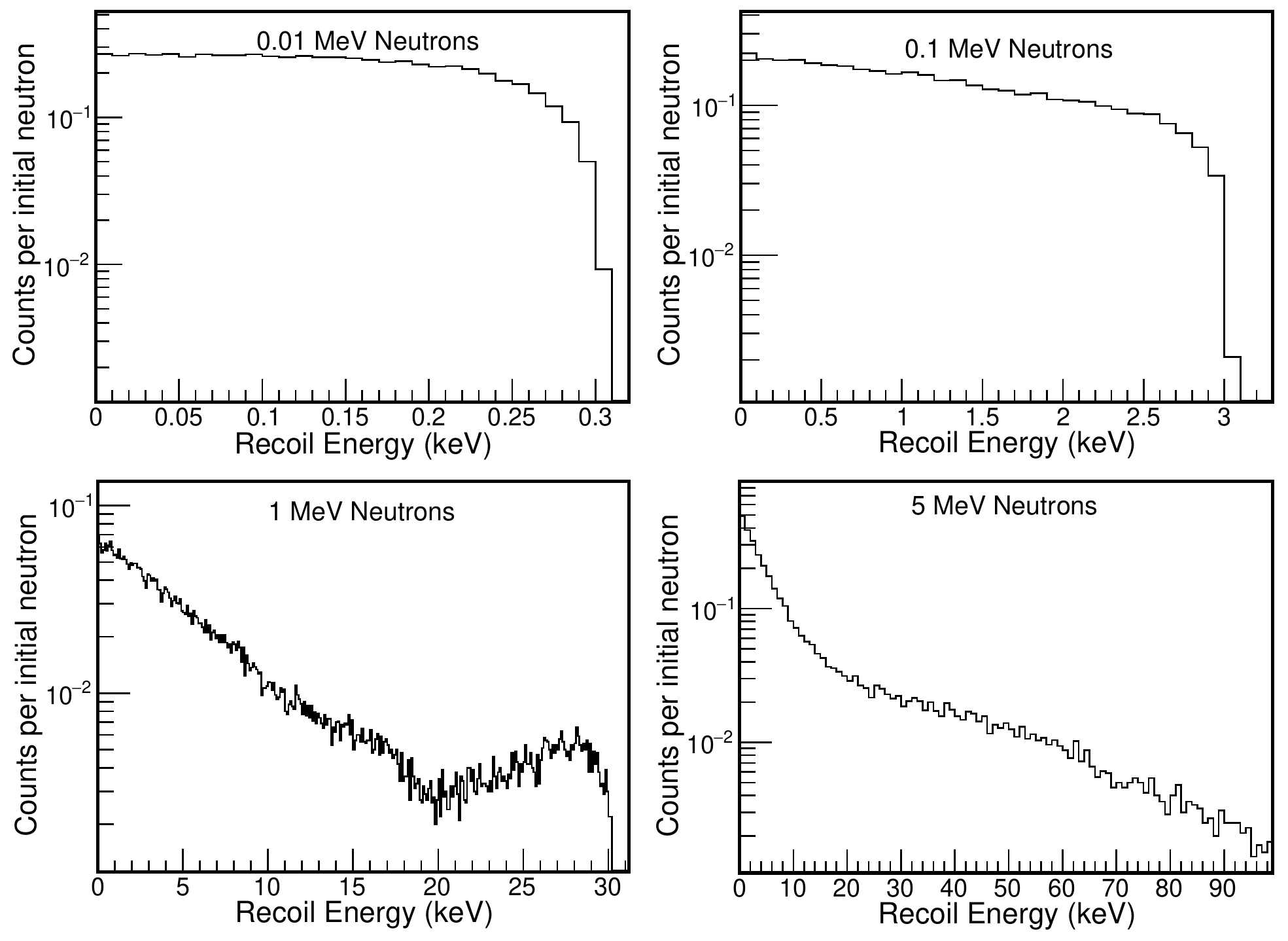}
\caption{$\Xe132$ recoil spectra generated by monoenergetic neutrons of 
different energies.} 
\label{fig:rspec_mono_panel}
\end{figure}

Although at the neutron energies of our interest the elastic scattering 
dominates~\cite{ENDF}, inelastic processes can also occur whereby, for 
example, the incident neutron is absorbed by a $\Xe132$ nucleus and  
re-emitted together with a gamma ray, resulting in a small recoil of the 
nucleus. The emitted neutron can then again scatter elastically or 
inelastically on another xenon nucleus depending on its energy, and so 
on. Each neutron is tracked until it hits and exits the outer boundary 
of the simulation volume or stops within the medium.  
Actually, given their large mean free paths, most neutrons in the energy 
range of the \nuein\  spectrum shown in Fig.~\ref{fig:n-spectrum-NO} 
would be scattered out of the detector volume after undergoing single or 
multiple scattering.  

To understand the nature of the final recoil spectrum, we first study 
the recoil spectra generated by initially monoenergetic neutrons. 
Fig.~\ref{fig:rspec_mono_panel} shows the xenon recoil spectra for 
four different initial energies of neutrons, namely, 0.01, 0.1, 1 and 5 
MeV.  
It is seen that while at very low energies the recoil spectrum is almost 
flat, the spectrum becomes increasingly skewed towards low recoil 
energies as the initial neutron energy increases. This directly reflects 
the behavior of the angular distribution of the recoiling nucleus in 
the elastic scattering of neutrons on $\Xe132$ nuclei, with relatively 
more number of forward (small angle) scattering taking place than 
backward scattering as the neutron energy increases~\cite{ENDF}. 
\begin{figure}
\includegraphics[width=0.9\columnwidth]{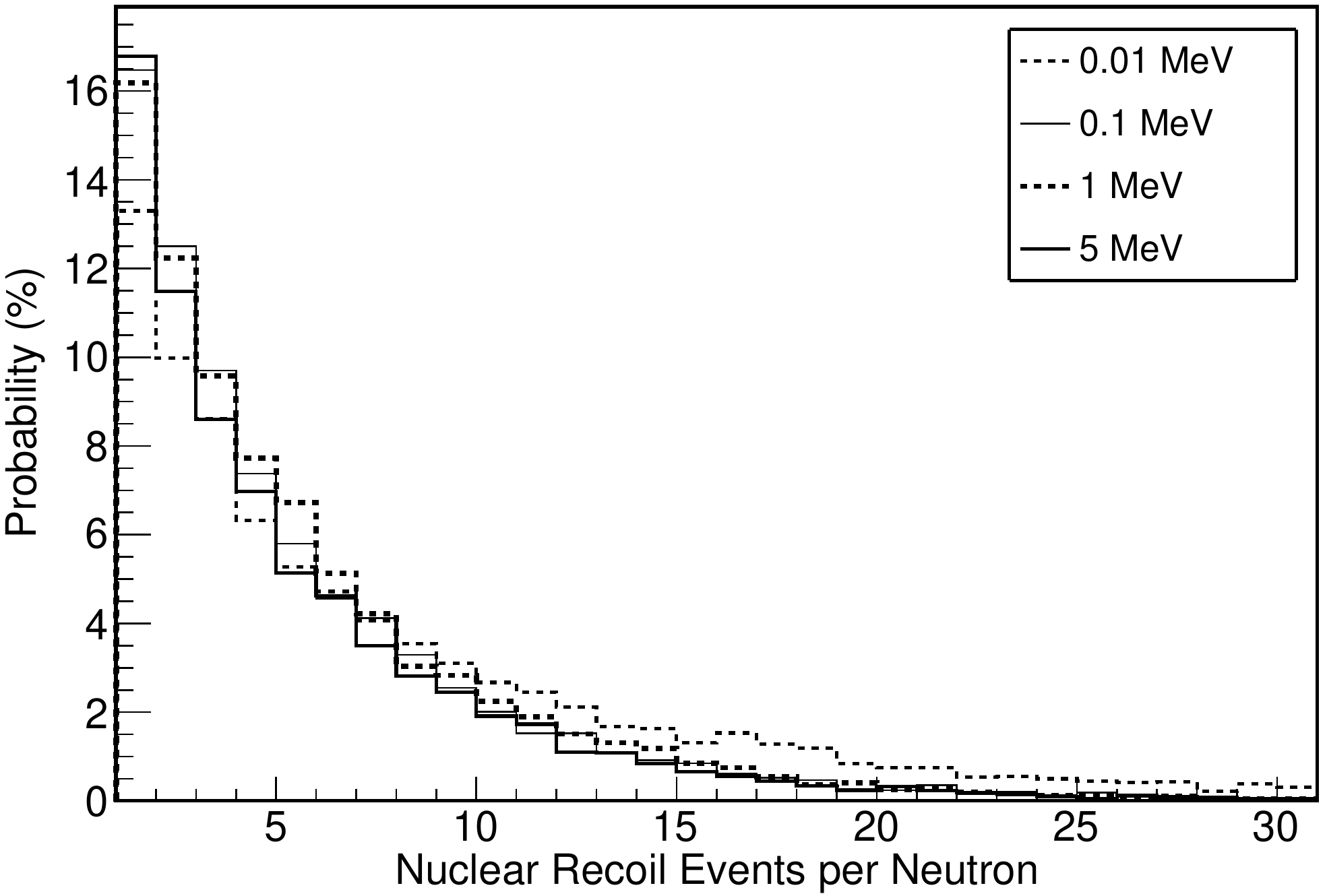}
\caption{Distribution of the number of $\Xe132$ recoils produced 
(i.e., the number of scatterings suffered) by 
monoenergetic neutrons of various energies.}    
\label{fig:recoil-number-dist}
\end{figure}

Next, we show in Fig.~\ref{fig:recoil-number-dist} 
the probability distribution of the number of $\Xe132$ recoils produced 
by neutrons of different initial energies. It can be seen that for all 
initial neutron energies, 1--2 scattering 
events dominate, with the probability of larger number of scattering  
falling off faster with number of scatterings for increasing 
initial energy of the neutron. Note, however, that the multiple 
scattering events (greater than 1--2 scatterings), although less 
probable than 1--2 scattering events, can still contribute significantly 
to the total recoil spectrum because each such scattering produces a 
larger number of recoils per initial neutron than the 1--2 scattering 
events. 

\begin{figure}
\includegraphics[width=0.9\columnwidth]{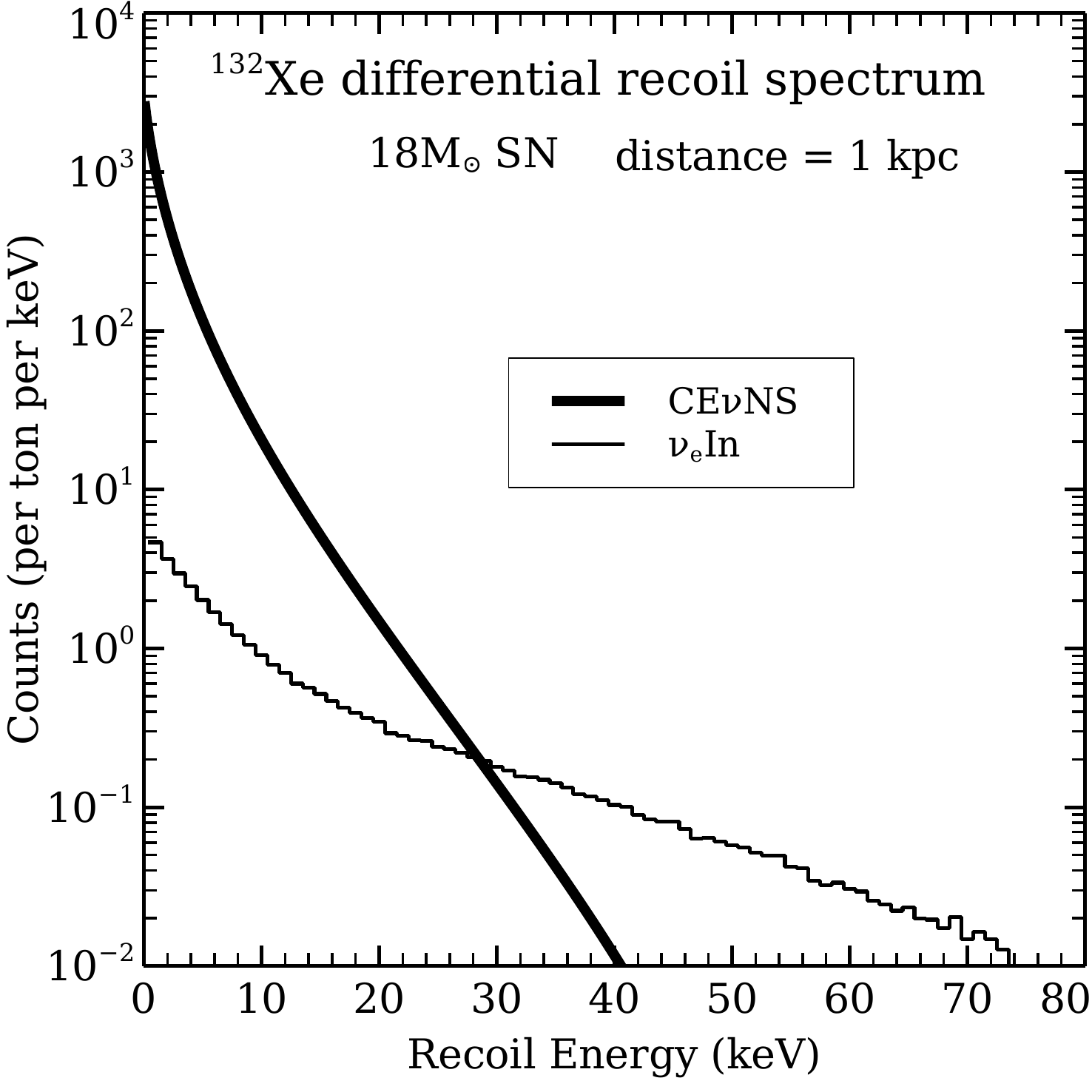}
\caption{$\Xe132$ differential recoil spectrum due to supernova 
neutrino induced \cenns\  (curve marked ``\cenns") and the spectrum due 
to neutrons produced by inelastic CC interaction of supernova $\nue$s 
with $\Xe132$ nuclei (curve marked ``\nuein") 
for the case of normal ordering of neutrino mass hierarchy shown in 
Fig.~\ref{fig:n-spectrum-NO}. Both the curves correspond to SN 
neutrino flux given by the 
Basel-Darmstadt simulations~\cite{BD-SN-sim-10} of a
$18\msun$ supernova progenitor at a distance of 1 kpc.}
\label{fig:cenns_nuin-Xe-diff-recoil-spect}
\end{figure}

The final differential recoil spectrum generated by the SN 
$\nue$ induced neutrons whose initial energies are randomly sampled from 
the spectrum shown in Fig.~\ref{fig:n-spectrum-NO} is shown in 
Fig.~\ref{fig:cenns_nuin-Xe-diff-recoil-spect} together with the 
direct recoil spectrum of $\Xe132$ due to the \cenns\ 
process given by Eq.~(\ref{eq:dNRdER}). 
From Fig.~\ref{fig:cenns_nuin-Xe-diff-recoil-spect} we see that 
the \nuein\  generated xenon recoils associated with the \nueCC\  
events can in principle give a significant contribution to the total 
observable NR signal at relatively large
($\gsim$ 30 keV) recoil energies where the \cenns\ 
generated NR signal contribution is relatively small.

\section{Observable signals in dual-phase xenon detectors}
\label{sec:S1-S2} 
A typical dual-phase (liquid-gas) xenon detector consists of a 
cylindrical time projection chamber (TPC) containing liquid xenon, with 
a layer of gaseous xenon at the top. The energy deposited due to a 
particle interaction within the TPC is dissipated in the liquid xenon 
medium through various modes including excitation and ionization of the 
xenon atoms, and heat dissipation. This 
gives rise to two observable signals: (a) a prompt scintillation light 
signal (S1) associated with deexcitation of the excited xenon molecular 
states and (b) a delayed light signal associated with the ionization 
electrons which are drifted through the liquid xenon 
by an applied vertical electric field and extracted (by a second, 
relatively stronger electric field) into the xenon gas layer at the 
top where they give rise to a secondary, ``proportional scintillation"  
signal (S2) (it being proportional to the number of extracted 
ionization electrons) by collision with xenon atoms. 
Both these light signals are detected with two arrays of photomultiplier 
tubes (PMTs), one placed at the top and the other at the bottom of the 
TPC. In terms of the number of photoelectrons (PEs) produced in the 
PMTs, the S2 signal is generally significantly larger than S1 (by 
factors ranging from a few tens to a few hundreds depending on the 
energy deposition) due to amplification characteristic of the 
proportional scintillation process. Detection of both S1 and S2 signals 
from the same energy deposition event gives a dual-phase TPC the ability 
to reconstruct the 3D location of the energy deposition event within the 
TPC, while the measured S1-to-S2 ratio allows effective discrimination 
between NR signals (due to dark matter, background neutrons, for 
example) and ER signals 
(due to background $\beta$ and $\gamma$ particles). 

Below we discuss the observable S1 and S2 
signals due to \cenns\   origin xenon nuclear recoils and \nueCC\  origin  
neutrons (which generate xenon nuclear recoils), electrons and  
$\gamma$-rays, in a generic dual-phase LXe detector. The 
contributions of the different signal components associated with the
\cenns\  and \nueCC\  events are assessed and the possibility of
distinguishing amongst them is discussed in
Sec.~\ref{sec:event-type-contributions} below. The basic physics 
we follow for working out these signals is that described in the Noble 
Element Simulation Technique (NEST) 
model~\cite{nest-paper-1,nest-paper-2,nest-paper-3}, which gives a 
semi-empirical method of estimating the yields of scintillation photons   
and ionization electrons produced by recoiling particles in LXe. Because 
of the basic difference in the nature of energy deposits 
by NRs and ERs, the scintillation and ionization yields differ in the 
two cases, and consequently, the S1 and S2 signal generation procedures 
are somewhat different in the two cases. In this paper we are interested 
in obtaining only the strengths of the signals in a generic 
detector and not detailed detector specific simulations. As such we 
follow the simplified treatment described in 
Refs.~\cite{lang-etal-16,Xenon1t-16} together with 
GEANT4 simulation (for propagation and tracking of the relevant 
particles inside the detector) for obtaining 
the S1 and S2 signal strengths. For the NRs, we use the same light and
ionization yields as those used in Ref.~\cite{LUX-2016} and detector 
parameters (like those specifying S1 and S2 gains, drift time of
the ionization electrons, and so on) as those used in 
Ref.~\cite{lang-etal-16}. For the ER case, we use the benchmark results 
for the light and ionization yields using the publicly available NEST 
code~\cite{NEST-code-link}, while the detector parameters are taken to 
be the same as those used for the NR case. More details of our 
simulation procedure are given in 
Ref.~\cite{Sayan-simu-paper-in-prep-2022}. 
\subsection{S1 and S2 signals from \cenns\  and \nueCC\  origin 
nuclear recoils} 
\label{subsec:S1-S2-cenns-nuin}
\begin{figure}[ht]
\centering
\begin{minipage}[b]{0.9\linewidth}
\includegraphics[width=0.9\columnwidth]{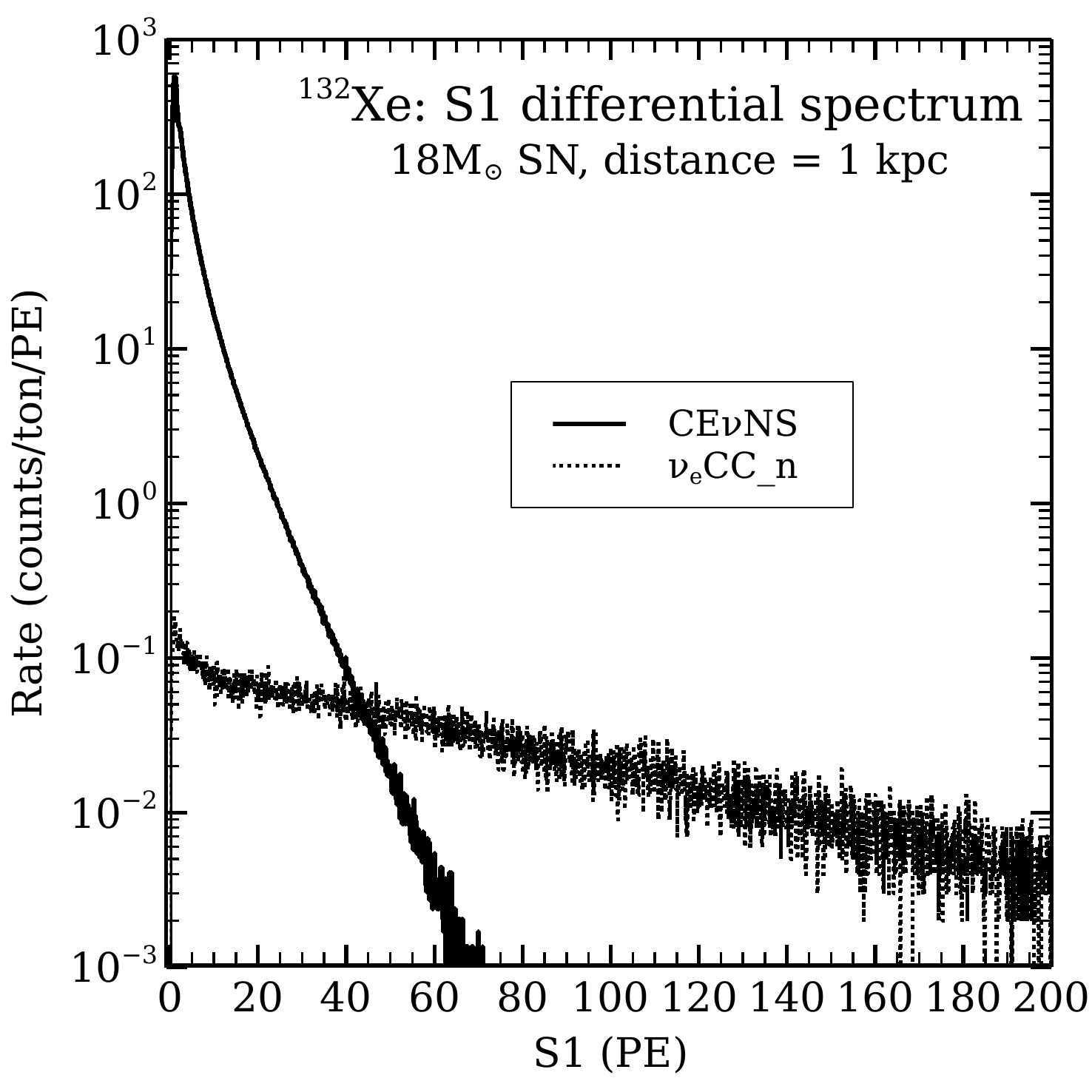}
\end{minipage}
\quad
\begin{minipage}[b]{0.9\linewidth}
\includegraphics[width=0.9\columnwidth]{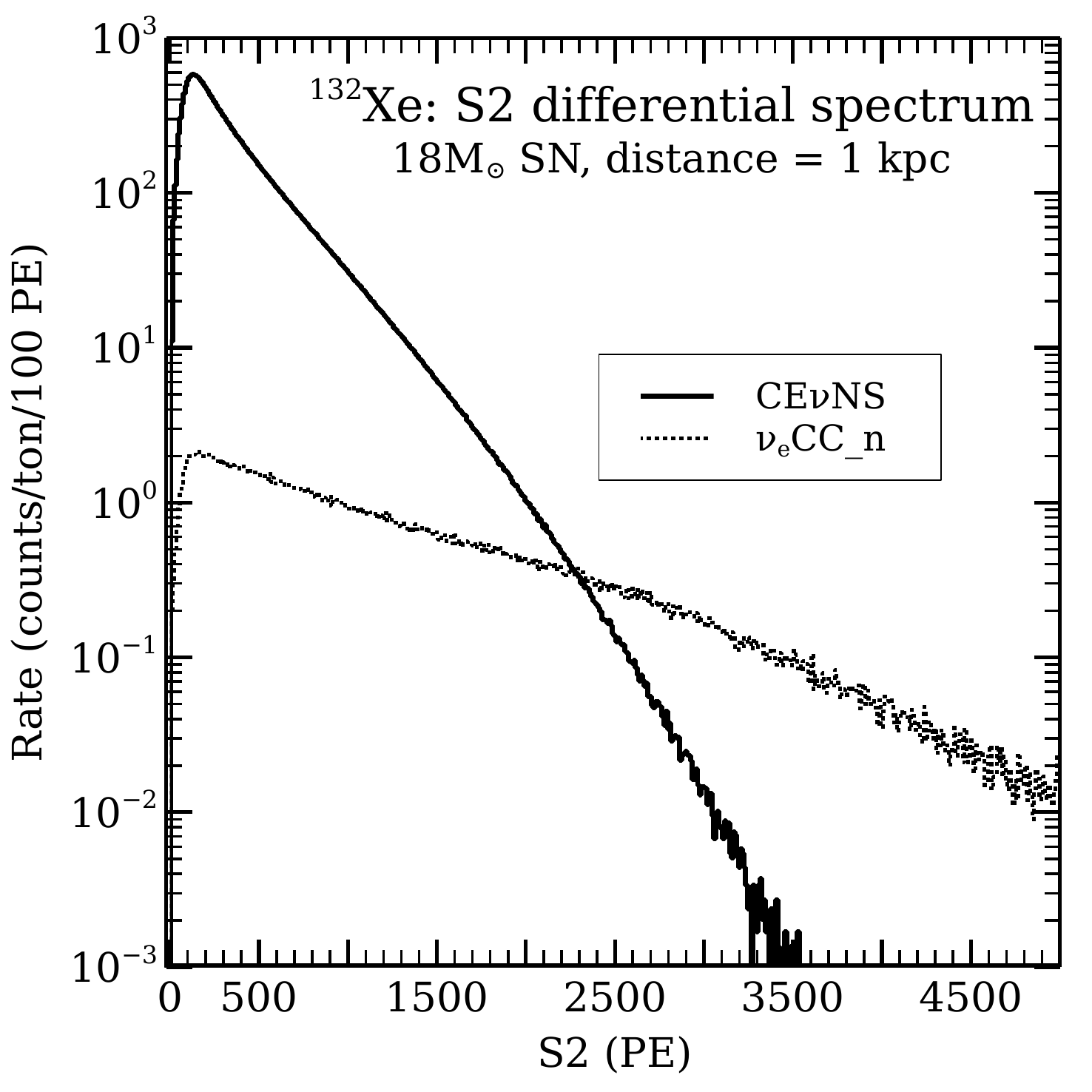}
\end{minipage}
\caption{The observable S1 (upper panel) and S2 (lower panel) 
differential spectra corresponding to the $\Xe132$ recoil spectra shown 
in Fig.~\ref{fig:cenns_nuin-Xe-diff-recoil-spect}.}
\label{fig:cenns_nuin_S1-S2_diff_spect}
\end{figure}

The S1 and S2 differential spectra due to \cenns\  and \nueCC\     
origin NRs are generated from the corresponding differential recoil 
energy spectrum shown in Fig.~\ref{fig:cenns_nuin-Xe-diff-recoil-spect}. 

For the \cenns\  process, a given incoming neutrino may be expected 
to interact, if at all, only once within the detector volume and thus 
generate only one xenon nuclear recoil, which gives one S1 and one S2 
signal for each \cenns\  interaction. On the other hand, a 
single neutron produced in the \nueCC\  process may undergo multiple 
scattering giving rise to multiple xenon recoils along its track, which 
would in general give rise to multiple S1 and S2 signals associated 
with a single neutron. However, due 
to different nature of their origin, the construction of the final   
observable S1 spectrum must be treated differently from that of S2. 

First, we note that the typical signal sampling rate in 
currently operating large LXe detectors is about 1 sample per 
10 ns, a time scale that is much larger than the light travel time scale 
across the detector volume. So, in the case of S1, it being a prompt 
signal, the arrival time difference (at the PMT arrays) 
between S1 photons originating from two successive interaction vertices 
of a neutron is essentially the same as the time interval between the 
two successive interactions. The typical width of an individual S1 pulse  
is about 27 ns~\cite{Aprile-Doke_10} (fixed by the mean 
deexcitation lifetime of the excited xenon atoms). So any two S1 signals 
associated with two neutron scattering events originating within say 50 
ns of each other will overlap and cannot be distinguished. A neutron of 
initial energy of $\sim$ 2 MeV (corresponding to the peak of the neutron 
spectrum shown 
in Fig.~\ref{fig:n-spectrum-NO}) has a speed of $\sim$ 2 cm/ns. With a 
mean free path of $\sim$ 13 cm in liquid xenon, the mean time between 
two successive scatterings of such a neutron is $\sim$ 6.5 ns. This 
implies that multiple S1 signals from neutron scattering events 
separated by less than 7--8 mean free paths would be unresolvable and 
will be merged. At the same time, as seen from 
Fig.~\ref{fig:recoil-number-dist}, the probability 
for a neutron undergoing more than 7--8 scatterings is 
relatively small unless the initial neutron energy is significantly 
smaller than 2 MeV and the detector scale size is significantly larger 
($>$ 1 m). Based on these considerations, in our simulation program the 
S1 signals (PEs) due to multiple scattering of each neutron 
(sampled from the spectrum shown in Fig.~\ref{fig:n-spectrum-NO}) are 
merged into one S1 signal. 

On the other hand, the S2 signals are produced by the ionization 
electrons colliding with the xenon atoms in the gas phase, the electrons 
being originally produced in the liquid xenon bulk of the TPC and then 
drifted in the vertical ($z$) direction through the liquid xenon and 
finally extracted into the gas phase by another external electric field. 
With a typical drift speed of the electrons of $\sim 
2\,{\rm mm}/\musec$ through liquid xenon, the arrival time difference
at the gas phase between two sets of ionization electrons originating 
from two successive neutron scattering events (occurring in 
the liquid xenon) can be $\sim$ a fraction 
of a $\musec$ to a few hundreds of $\musec$ (depending on the 
difference in the $z$ positions of the two neutron scattering vertices 
in the liquid and the dimensions of the detector), compared to a few 
ns time scale between two successive scattering of a neutron in the 
liquid xenon. Given that observed S2 signal pulses have typical widths 
of $\sim$ 1--2 $\musec$~\cite{Xenon1t-19}, in our simulation, 
therefore, the S2 signals (PEs) due to electrons arriving in the gas 
phase within a $2\musec$ window were combined into a single S2 
signal, while those due to the electrons reaching the gas phase beyond 
this window were treated as separate ones.
The resulting S1 and S2 differential spectra due to $\Xe132$ recoils  
originating from \cenns\ and \nuein\  processes  
(Fig.~\ref{fig:cenns_nuin-Xe-diff-recoil-spect}) are shown in 
Fig.~\ref{fig:cenns_nuin_S1-S2_diff_spect}.

\subsection{S1 and S2 signals from electrons and deexcitation gamma 
rays}
\label{subsec:S1-S2-egamma}
\begin{figure}[ht]
\centering
\begin{minipage}[b]{0.9\linewidth}
\includegraphics[width=0.9\columnwidth]{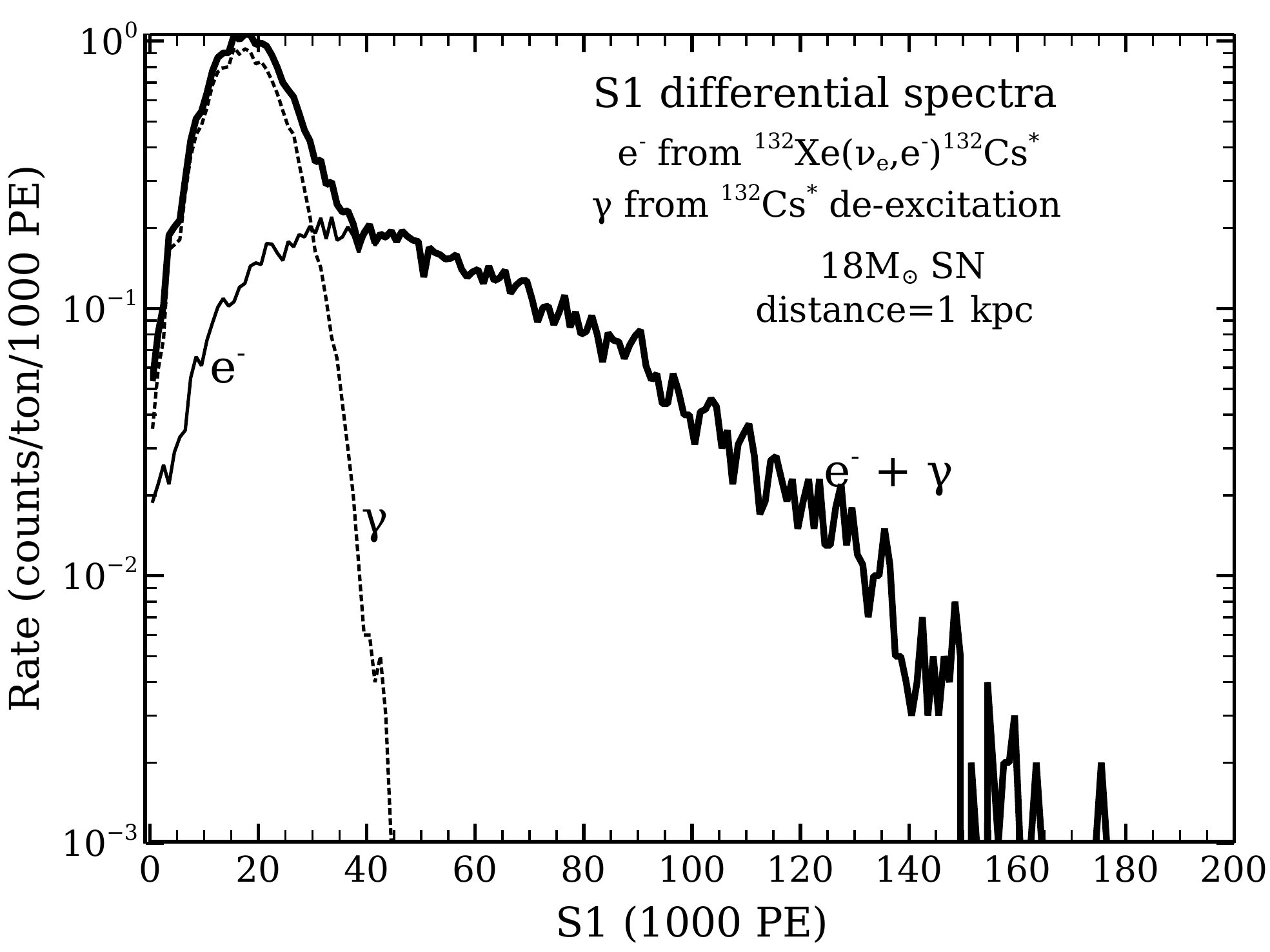}
\end{minipage}
\quad
\begin{minipage}[b]{0.9\linewidth}
\includegraphics[width=0.9\columnwidth]{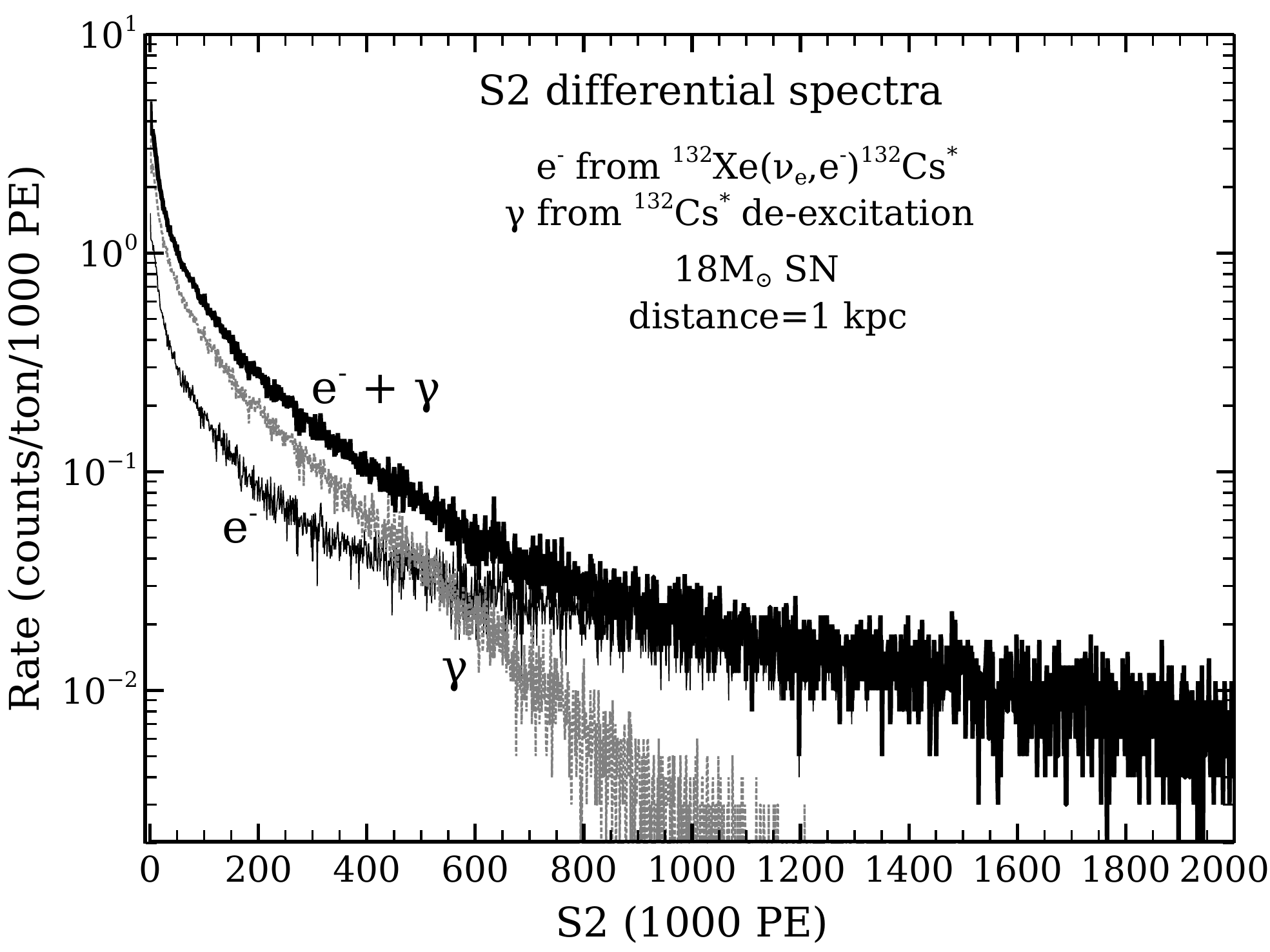}
\end{minipage}
\caption{The observable S1 (upper panel) and S2 (lower panel) 
differential spectra due to electrons generated in the CC process 
$\Xe132(\nue\,, \eminus){\Cs132}^*$ and $\gamma$-rays 
from deexcitation of the ${\Cs132}^*$ nuclei, corresponding to their 
energy spectra shown in Figs.~\ref{fig:e-spectrum-NO} 
and \ref{fig:gamma-spectrum-NO}, respectively.} 
\label{fig:egamma_S1-S2_diff_spect}
\end{figure}

The energy spectra of the electrons produced in the primary CC 
process $\Xe132(\nue\,, \eminus){\Cs132}^*$ and the $\gamma$-rays 
from deexcitation of the ${\Cs132}^*$ nuclei are shown in 
Figs.~\ref{fig:e-spectrum-NO}  and \ref{fig:gamma-spectrum-NO}, 
respectively. While the low energy electrons create further ionization 
electrons, the high energy ones create Bremstrahlung $\gamma$-rays, 
which together with the original $\gamma$-rays undergo Compton and 
photoelectric interactions with electrons in the xenon atoms and produce 
further energetic electrons. We 
perform GEANT4 simulations to propagate and track all the electrons 
and gamma rays whose initial energies are 
randomly sampled from their energy spectra shown in 
Figs.~\ref{fig:e-spectrum-NO} and \ref{fig:gamma-spectrum-NO}, 
respectively, and then generate their S1 and S2 signals by the procedure 
outlined above. It is observed that although each electron or  
$\gamma$-ray undergoes many interactions, all these interactions occur 
within a small region (within a few mm to few cm depending on the energy 
of the particle) around their production 
points. For reasons explained in 
Sec.~\ref{subsec:S1-S2-cenns-nuin}, this effectively gives rise 
to one S1 signal and about 1--3 S2 signals (after taking into account 
the effect of merging of S2 signals) for every electron or $\gamma$-ray. 

The resulting S1 and S2 differential spectra generated by the electrons 
and $\gamma$-rays are shown in Fig.~\ref{fig:egamma_S1-S2_diff_spect}

\section{Contributions of different event types and their 
distinguishability in the observed signals} 
\label{sec:event-type-contributions} 
In a SN burst event, a LXe detector would in general record both 
the \cenns\  events due to NC interaction of all species of SN 
neutrinos with the xenon nuclei and the \nueCC\  events due 
to CC interaction of the $\nue$s with the xenon nuclei. The \cenns\  
events generate signals characteristic of nuclear recoils  
whereas the \nueCC\   events in general have two signal components: (a) 
the ``\nueCCegamma"\  component due to electrons 
generated in the CC process $\Xe132(\nue\,, \eminus){\Cs132}^*$ and 
additionally due to $\gamma$-rays from deexcitation of the 
final state ${\Cs132}^*$ nuclei, and (b) the ``\nueCCn"\  component due 
to xenon nuclear recoils caused by the scattering of neutrons off the 
xenon nuclei, the neutrons being from deexcitation of the final state 
${\Cs132}^*$ nuclei. 

\begin{figure}[ht]
\centering
\begin{minipage}[b]{0.9\linewidth}
\includegraphics[width=0.9\columnwidth]{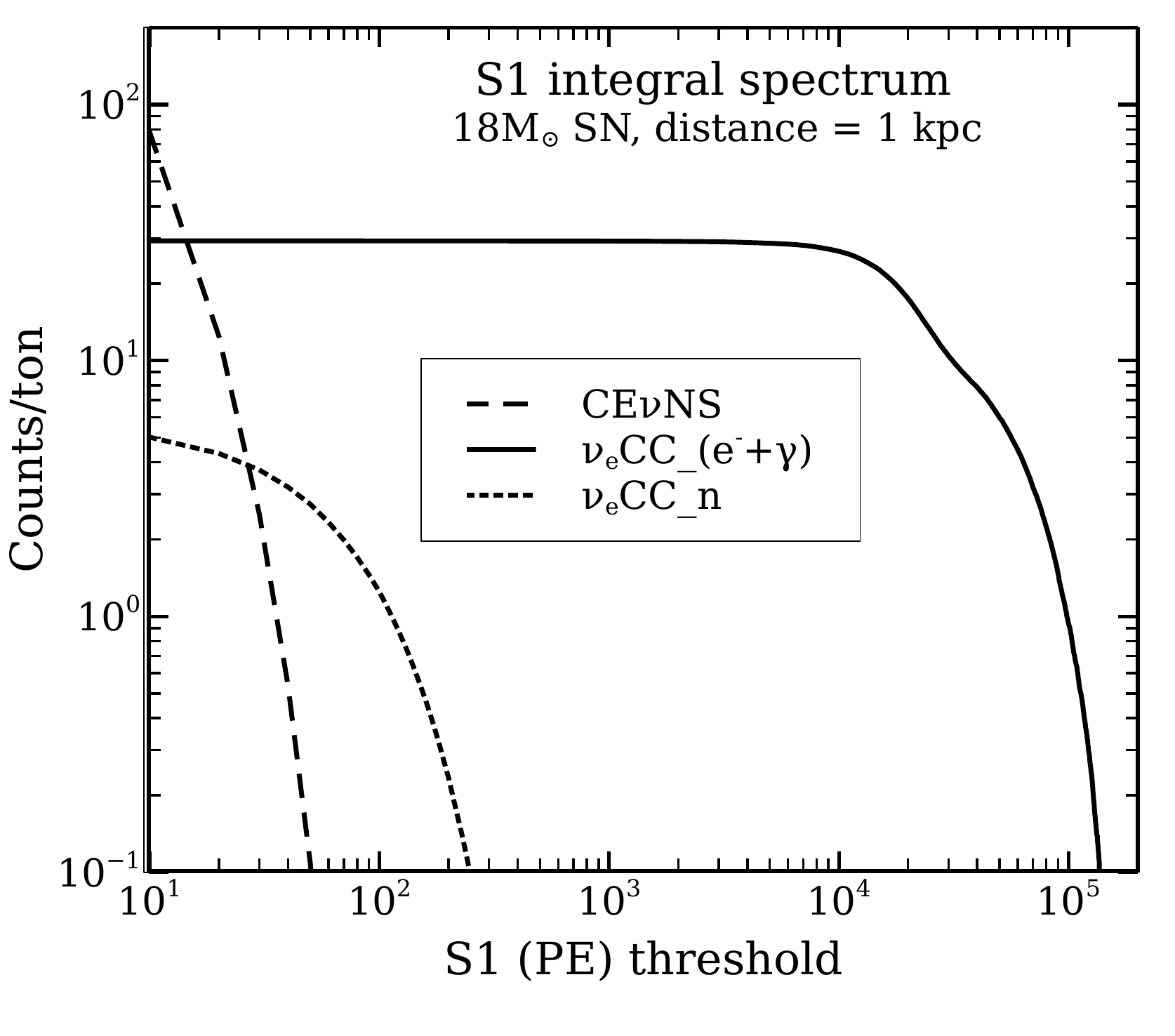}
\end{minipage}
\quad
\begin{minipage}[b]{0.9\linewidth}
\includegraphics[width=0.9\columnwidth]{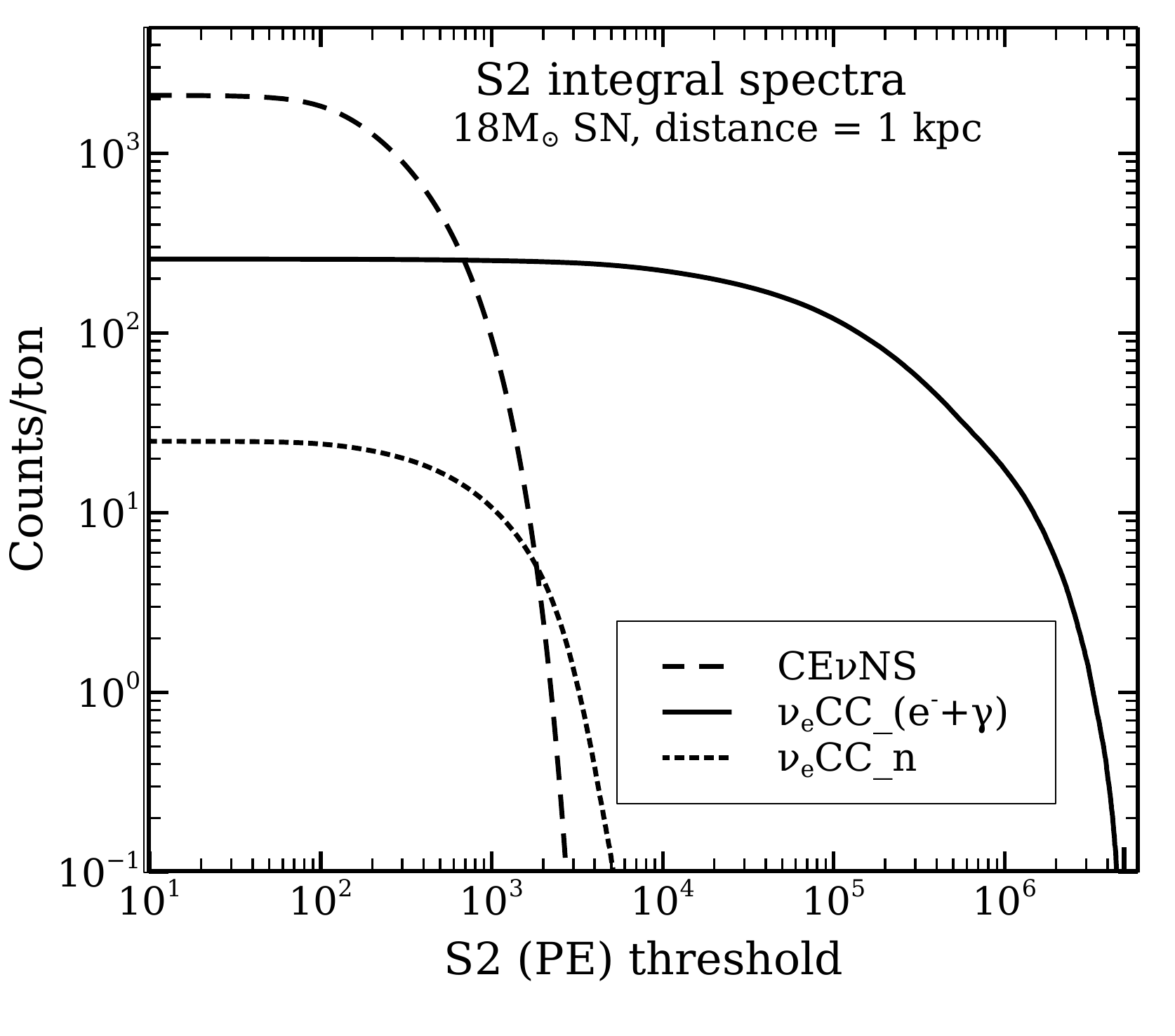}
\end{minipage}
\caption{Contributions of different signal components to the 
observable S1 (upper panel) and S2 (lower panel) signals as functions of 
the S1 (PE) and S2 (PE) threshold, respectively, per ton of liquid 
$\Xe132$. These are obtained by respectively integrating the 
differential S1 and S2 spectra 
shown in Figs.~\ref{fig:cenns_nuin_S1-S2_diff_spect} and 
\ref{fig:egamma_S1-S2_diff_spect} for different 
values of S1 and S2 thresholds.}
\label{fig:S1-S2_int_cenns_nuin_egamma}
\end{figure}

In order to see the contribution of different event types in the 
final observed number of events, we show in 
Fig.~\ref{fig:S1-S2_int_cenns_nuin_egamma} the number of 
events as functions of S1 and S2 PE thresholds obtained by 
respectively integrating the differential S1 and S2 spectra shown in 
Figs.~\ref{fig:cenns_nuin_S1-S2_diff_spect} and 
\ref{fig:egamma_S1-S2_diff_spect} for different values 
of S1 and S2 thresholds, respectively.  

Note that S2 signals are in general significantly larger than S1 
signals by at least an order of magnitude both in terms of number of 
photoelectrons in individual signal events as well as in terms of 
integral number of events (counts/ton) as function of the relevant PE 
threshold. 

It is clear that \nueCCegamma\   signals largely dominate the overall 
observed signals in both S1 and S2 except for relatively low detector 
thresholds with 
S1 threshold, $\Soneth\lsim$ 15, and S2 threshold  
$\Stwoth\lsim$ 700 PEs, for which \cenns\  origin NR signals dominate. 
This indicates that upcoming large, multi-ton class LXe based DM 
detectors will be strongly 
sensitive to the $e^{-}+\gamma$ signal arising from \nueCC\  
events caused by SN $\nue$s in addition to the nuclear recoil signal 
due to the \cenns\   events arising from NC interactions of all the six 
neutrino species. 
The usual problem of ER and NR backgrounds encountered in DM search may 
be largely mitigated in the case of SN neutrinos due to 
relatively short duration ($\sim O(10)$s) of the SN burst event. 
Note that the \nuein-induced NR signal (\nueCCn) is subdominant to both 
\cenns-induced NR and \nueCCegamma\   signals.  

From the estimated numbers of various types of signals shown in  
Fig.~\ref{fig:S1-S2_int_cenns_nuin_egamma}, and nominally scaling these 
numbers with active detector mass ($M_{\rm Det}$) and 
distance ($d$) to the SN as $M_{\rm Det}/d^2$, we see that with a future 
large LXe detector of the class of the proposed 
DARWIN~\cite{multi-ton-dm-detectors} detector, for example, 
with an active target mass of $\sim$ 40 ton, we may expect $\sim$ 
[839, 103, 10] S2 signals due to [\cenns, \nueCCegamma,   
\nueCCn] above S2 thresholds of 10 PEs, for the 
$18\msun$ SN at a distance of 10 kpc. For a S2 threshold of 
$\Stwoth=100\,$ and 1000 PEs, these numbers are $\sim$ [730, 103, 9] and 
$\sim$ [37, 101, 4], respectively. Similar scaling of expected 
number of 
signals with detector mass and SN distance can be obtained for S2 as 
well as S1 for various S2 and S1 PE thresholds.

An important question to consider is whether 
it would be possible to distinguish the individual signal components 
(\cenns, \nueCCegamma, \nueCCn) within a given total number of 
observed events. This is important because, while the number of 
\cenns\   events is sensitive to all six neutrino species, those of 
the two signal components of \nueCC\  events (\nueCCegamma\  and 
\nueCCn) are both sensitive only to $\nue$s. Identification of each 
signal component in the total number of observed events would therefore 
provide a good handle on the flavor composition of the total SN neutrino 
flux. 

\begin{figure}[ht]
\centering
\begin{minipage}[b]{0.9\linewidth}
\includegraphics[width=0.9\columnwidth]{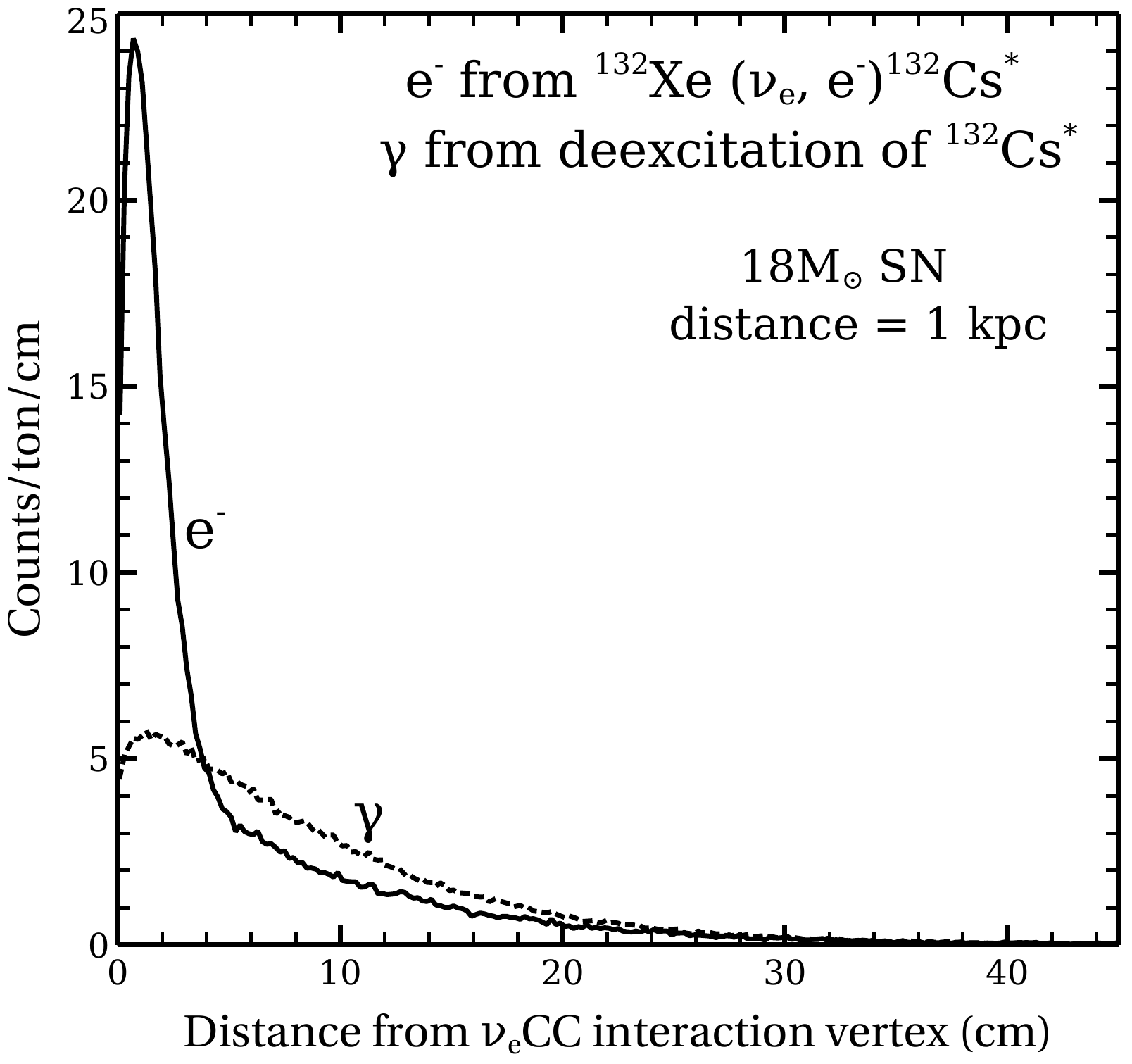}
\end{minipage}
\quad
\begin{minipage}[b]{0.9\linewidth}
\includegraphics[width=0.9\columnwidth]{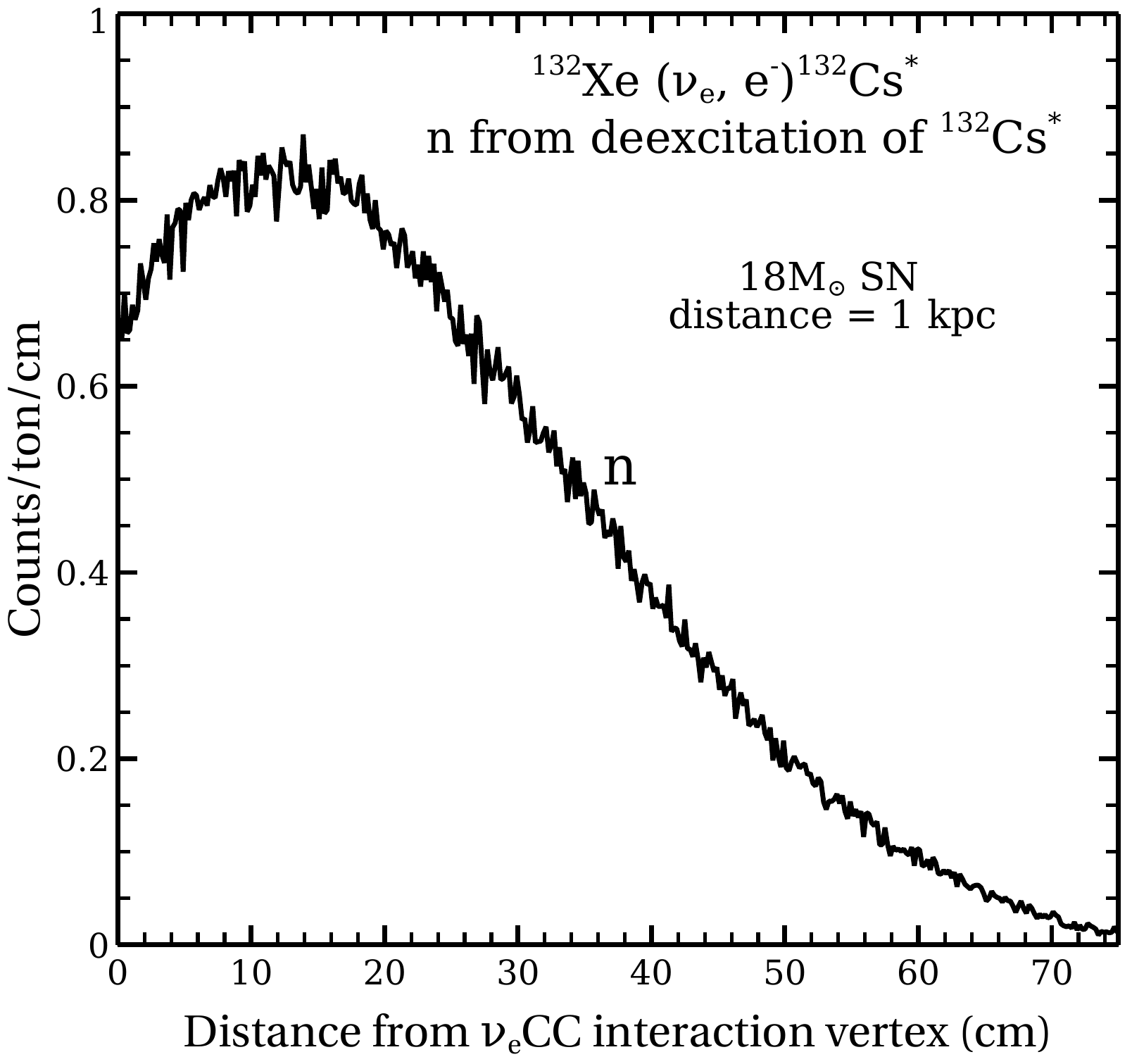}
\end{minipage}
\caption{Hit-point distribution for electrons and gamma rays (upper 
panel) and neutrino-induced neutrons (lower panel) from \nueCC\  
interactions.}
\label{fig:hit-points_e-gamma-n}
\end{figure}

In this context, note first that, since the \nueCC\  and the 
\cenns\   events result from different underlying fundamental 
interaction processes, the occurrence of these two types of events 
would in general be separated in time. On 
the other hand, for a given \nueCC\  interaction event, the prompt S1 
signals due to \nueCCegamma\  and \nueCCn\   will be 
temporally superimposed on each other, and will be registered as a 
single combined S1 response, thus rendering them indistinguishable. 
However, the following considerations with regard to the S2 signals may 
still offer a possible way to separate the \nueCCegamma\  and \nueCCn\  
signals, at least on a 
statistical basis, though perhaps not on event-by-event basis: 

In general, a particle (electron, gamma ray or neutron) produced in a 
\nueCC\   interaction event will undergo multiple scattering whereby the 
particle would deposit a part of its energy (through production of 
a bunch of ionization electrons which eventually produce the S2 
signals) at each ``hit-point", i.e., a point of interaction with a 
scatterer (e.g., a xenon nucleus or an electron) along its track. Since 
the 
ranges of the electrons and $\gamma$-rays of the relevant energies are 
of the order of a few mm to a few cm in LXe~\cite{e-gamma-range}, the 
majority of these ``hit-points" would be located very close (within a 
few cm) to their corresponding \nueCC\   interaction points where the 
electrons and gamma rays are originally generated. In contrast, 
the neutrons mainly lose energy through multiple 
scatterings on the xenon nuclei, thereby producing recoiling xenon 
nuclei which, in turn, produce the ionization electrons. Since the mean 
free path lengths of the neutrons in LXe is relatively large ($O(10)$ 
cm), it is expected that the hit-points for neutrons would in general be 
located further away (compared to those for electrons and gamma rays) 
from the corresponding \nueCC\  interaction points where the neutrons 
are generated. This can be seen from 
Fig.~\ref{fig:hit-points_e-gamma-n} which shows the distribution of all 
the hit points (marked by their distances from the corresponding \nueCC\   
interaction vertices) for neutrons, electrons and gamma rays, obtained 
from our GEANT4 simulation.  
Clearly, the hit-point distribution of the neutrons is broader, and the 
neutrons' energy deposition points are spatially more widely spread out,  
than those for electrons and gamma rays. Since the LXe TPCs generally 
have the capability of hit-point position 
reconstruction using the S2 signals from the arrays of PMTs above the 
gas phase of the TPC, a suitable algorithm similar to 
the one used for (x,y) position reconstruction of the hit-points (as 
discussed, for example, in Sec.~II.B of Ref.~\cite{Xenon1t-19-PRD}) 
could in principle be used to distinguish between the \nueCCegamma\   
and \nueCCn\   signals, at least on a statistical basis, provided that 
corresponding S2 signals of the two event subclasses are temporally 
separable, i.e., the corresponding 
track projections are well separated along the TPC drift ($z$) axis.  
Indeed, from a simple integration of the distance distributions shown in 
Fig.~\ref{fig:hit-points_e-gamma-n} for various threshold distances, one 
finds, for example, that more than 75\% of the neutron hit-points lie 
beyond a distance of 10 cm from the corresponding \nueCC\  interaction 
vertices, whereas the fractions of gamma ray and electron hit points 
that lie beyond 10 cm from the corresponding \nueCC\   interaction 
vertices are only $\sim$ 17\% and 32\%, respectively, with the rest 
(83\% and 68\%, respectively) of the hit points lying inside 10 cm. Of 
course, to see how exactly these different hit-point 
distributions at the energy deposition level in the bulk of the 
TPC translate to different spatial distributions of the PMT hit patterns 
at the PMT array plane at the top of the gas phase, will require 
detailed detector specific simulations, which is beyond the scope of the 
present paper.        

At this point it is worth mentioning that there is also an additional 
channel for possible identification of the neutrons generated in the 
\nueCC\  interactions. This comes from the consideration that a large 
fraction of the neutrons associated with the \nueCC\   events   
may not be stopped within the LXe tank, and may escape the 
detector volume with residual energies. Indeed, our neutron tracking 
simulation described in Sec.~\ref{subsec:geant4} shows this escaping 
neutron fraction to be more than 90\%. An array of suitable neutron 
detectors placed around the detector volume (see.e.g., 
Ref.~\cite{n-veto-detector-XENON-2020}) may, therefore, be utilized to 
detect the escaping neutrons in coincidence, thereby identifying the 
\nueCC\  origin neutrons. Again, the relevant neutron 
background is expected to be small over the duration of the SN burst 
event. 

Both the possibilities mentioned above are, however, challenging 
propositions that can be fully addressed and established only by means 
of detailed simulations using, for example, an ``event generator" 
similar to those used in neutrino physics (see, e.g., 
Ref.~\cite{MARLEY}), together with detailed detector simulations, which 
can be the subject of a future work. 
\section{Summary and conclusions} 
\label{sec:summary} 
\noindent 
Several previous studies have shown that future large, multi-10 tonne 
class, LXe detectors for dark matter search will also be sensitive to 
supernova neutrino induced xenon nuclear recoils due to the \cenns\  
process, which involves NC interactions of all six species of 
neutrinos including their antineutrinos. In this paper we 
have pointed out that the same LXe based detectors will also be 
sensitive to inelastic charged current (CC) interactions of the SN 
electron neutrinos with the xenon nuclei. In such interactions, an 
electron is produced in the final state and the post-interaction 
target nucleus is left in an excited state, the subsequent deexcitation 
of which produces, among other particles, gamma rays and neutrons. 
The electron and deexcitation gamma rays would give rise to 
electron recoil type signals (\nueCCegamma) while neutrons from the 
deexcitation of the post-interaction nuclei---the so called neutrino 
induced neutrons---would produce additional xenon nuclear recoils (i.e., in 
addition to \cenns\  generated NRs) through (multiple) scattering of the 
neutrons with the xenon nuclei and produce NR type signals (\nueCCn).  

We have estimated the relative contributions of the three different 
signal components (\cenns, \nueCCegamma, \nueCCn) to the total 
observable scintillation and ionization signals in a generic two-phase 
LXe detector. We have also discussed possible ways to identify the 
individual event types in the total number of observed events.   

It is found that in terms of the number of PEs produced in individual 
PMTs, the largest signals---both scintillation (S1) and ionization 
(S2) signals---come from the combined contribution due to the 
electrons generated in the \nueCC\   process
$\Xe132(\nue\,, \eminus){\Cs132}^*$ and the $\gamma$-rays
from deexcitation of the ${\Cs132}^*$ nuclei. In terms of event 
counts (number of events per tonne of active detector mass), the \cenns\  
events dominate the total event counts for relatively low detector 
thresholds of $\Soneth\lsim$ 15 and 
$\Stwoth\lsim$ 700 PEs, whereas the \nueCCegamma\  component of the 
\nueCC\   events dominate for larger thresholds. The \nueCCn\   
component of the \nueCC\   event signals remains subdominant 
to both the \nueCCegamma\   and \cenns\  signals throughout 
the S1 and S2 PE ranges under consideration. Moreover, in general, the 
prompt S1 signals corresponding to the \nueCCegamma\  and \nueCCn\  
signal components of any given \nueCC\   event will temporally be 
superimposed on each other, 
rendering them indistinguishable if only S1 signal is considered. 
However, it may still be possible to separate these two signal 
components, at least on a statistical basis, based on the expected  
different spatial distributions of the PMT hit patterns for the S2 
signals from \nueCCegamma\  and \nueCCn. In addition, since a large 
fraction of the neutrons generated in the \nueCC\  events are likely to 
escape the detector volume with significant amount of residual energies, 
coincidence detection of these escaping neutrons by means of a suitable 
arrangement of neutron veto detectors may provide an additional channel 
for identifying the \nueCC\  origin neutrons. Detection and 
identification of the \nueCC\   events due to 
SN $\nue$ CC interactions together with the \cenns\   events generated 
through NC interaction of all neutrino species  
in future multi-10 tonne class LXe detectors may provide a good probe 
for extracting useful information about the distribution of the total SN 
explosion energy going into different neutrino flavors.

{\bf Acknowledgment:} We thank the three anonymous referees for useful 
comments and constructive suggestions for improvement of the manuscript.  
We thank Rafael Lang for useful suggestions. One of us (SG) thanks Will 
Taylor for useful correspondence. Two of us (PB and SS) acknowledge 
support for this work under the Raja Ramanna Fellowship program of the 
Department of Atomic Energy (DAE), Government of India. 


\end{document}